\documentclass[12pt]{article}
\usepackage{graphicx}
\usepackage{axodraw}
\begin{document}
\baselineskip 0.6cm

\def\simgt{\mathrel{\lower2.5pt\vbox{\lineskip=0pt\baselineskip=0pt
           \hbox{$>$}\hbox{$\sim$}}}}
\def\simlt{\mathrel{\lower2.5pt\vbox{\lineskip=0pt\baselineskip=0pt
           \hbox{$<$}\hbox{$\sim$}}}}

\begin{titlepage}

\begin{flushright}
UCB-PTH-09/18
\end{flushright}

\vskip 2.0cm

\begin{center}

{\Large \bf 
Cosmic Signals from the Hidden Sector
}

\vskip 1.0cm

{\large Jeremy Mardon, Yasunori Nomura, and Jesse Thaler}

\vskip 0.4cm

{\it Center for Theoretical Physics, Department of Physics, \\
     University of California, Berkeley, CA 94720} \\
{\it and} \\
{\it Theoretical Physics Group, Lawrence Berkeley National Laboratory,
     Berkeley, CA 94720} \\

\abstract{Cosmologically long-lived, composite states arise as natural 
 dark matter candidates in theories with a strongly interacting hidden 
 sector at a scale of $10~\mbox{--}~100~{\rm TeV}$.  Light axion-like 
 states, with masses in the $1~{\rm MeV}~\mbox{--}~10~{\rm GeV}$ range, 
 are also generic, and can decay via Higgs couplings to light standard 
 model particles.  Such a scenario is well motivated in the context of 
 very low energy supersymmetry breaking, where ubiquitous cosmological 
 problems associated with the gravitino are avoided.  We investigate 
 the astrophysical and collider signatures of this scenario, assuming 
 that dark matter decays into the axion-like states via dimension six 
 operators, and we present an illustrative model exhibiting these 
 features.  We conclude that the recent data from PAMELA, FERMI, 
 and H.E.S.S. points to this setup as a compelling paradigm for 
 dark matter.  This has important implications for future diffuse 
 gamma ray measurements and collider searches.}

\end{center}
\end{titlepage}

\setcounter{tocdepth}{2}
\tableofcontents

\section{Introduction and Summary}
\label{sec:intro}

Weak scale supersymmetry is a very attractive candidate for physics beyond 
the standard model.  It stabilizes the weak scale against potentially 
large radiative corrections, leads to successful gauge coupling unification, 
and predicts a plethora of new particles accessible at the LHC.  This 
framework, however, also suffers from several generic cosmological problems, 
associated with overproduction of gravitinos or late decay of the field 
responsible for supersymmetry breaking~\cite{Weinberg:1982zq,Coughlan:1983ci,%
Moroi:1993mb,Kawasaki:2006gs}.  In many supersymmetry breaking scenarios, 
this requires a rather low reheating temperature after inflation, making 
it difficult to explain the observed baryon asymmetry of the universe.

A simple way to avoid these cosmological problems is to assume that the 
gravitino is very light.  If the gravitino mass satisfies
\begin{equation}
  m_{3/2} \simlt O(10~{\rm eV}),
\label{eq:m_32}
\end{equation}
then the gravitino is in thermal equilibrium with the standard 
model down to the weak scale.  The resulting gravitino abundance 
is small~\cite{Pagels:1981ke} and consistent with structure 
formation~\cite{Viel:2005qj}.  This allows for an arbitrarily 
high reheating temperature, and thus baryogenesis at high energies 
such as thermal leptogenesis~\cite{Fukugita:1986hr}.

It is remarkable that this simple cosmological picture is precisely 
the one suggested by arguably the simplest scheme for supersymmetry 
breaking.  Suppose that supersymmetry is dynamically broken at a scale
\begin{equation}
  \sqrt{F} \approx O(10~\mbox{--}~100~{\rm TeV}),
\label{eq:sqrt-F}
\end{equation}
giving $m_{3/2} = F/\sqrt{3}M_{\rm Pl} \approx O(0.1~\mbox{--}~10~{\rm eV})$, 
where $M_{\rm Pl} \simeq 2.4 \times 10^{18}~{\rm GeV}$ is the reduced 
Planck scale.  Then, if the sector breaking supersymmetry is charged 
under the standard model gauge group, gaugino and scalar masses of order 
$(g^2/16\pi^2) \sqrt{F} \approx O(100~{\rm GeV}~\mbox{--}~1~{\rm TeV})$ 
can be generated through standard model gauge loops~\cite{Dine:1981gu,%
Dine:1994vc}.  The generated squark and slepton masses are flavor 
universal, and thus solve the supersymmetric flavor problem.  The 
supersymmetric Higgs mass ($\mu$ term) can also be generated through 
direct interactions between the Higgs fields and the supersymmetry 
breaking sector~\cite{Csaki:2008sr,Komargodski:2008ax} or by the vacuum 
expectation value of a singlet field~\cite{Nilles:1982dy}.

What are the experimental signatures of this simple supersymmetry breaking 
scenario beyond its indirect implications on the superparticle spectrum? 
In this paper we advocate that the scenario may lead to distinct 
astrophysical and collider signatures, due to the following features 
of the dynamical supersymmetry breaking sector that may appear under 
rather generic conditions:
\begin{description}
\item{\bf Quasi-stable states:}
One immediate consequence of the present framework is that dark matter 
cannot be the lightest supersymmetric particle, which is the very light 
gravitino.  Dark matter, however, can arise naturally as a (quasi-)stable 
state in the supersymmetry breaking sector~\cite{Dimopoulos:1996gy}. 
Let $m_{\rm DM}$ be the mass of this state.  The annihilation cross 
section is then naturally $\langle \sigma v \rangle \approx (1/8\pi) 
(\kappa^4/m_{\rm DM}^2)$, where $\kappa$ represents typical couplings 
between states in the strong sector.  For $m_{\rm DM} \approx 
O(10~\mbox{--}~100~{\rm TeV})$, natural values for the coupling 
$\kappa \approx O(3~\mbox{--}~10)$ give $\langle \sigma v \rangle 
\approx (1/8\pi)(1/\mbox{TeV}^2)$, which leads to the correct thermal 
abundance for dark matter, $\Omega_{\rm DM} \simeq 0.2$.

In general, a strongly interacting sector of a quantum field theory often 
possesses enhanced global symmetries, such as baryon number and flavor 
symmetries.  These symmetries can lead to ``stable'' states if they 
are not broken by the strong dynamics.  It is, however, quite possible 
that these symmetries are not respected by physics at (much) higher 
energies, such as at the gravitational scale.  The ``stable'' states 
then become quasi-stable states, decaying through higher dimension 
operators on cosmological timescales.
\item{\bf Light axion-like states:}
A light state appears as a pseudo Nambu-Goldstone boson when an approximate 
global symmetry is spontaneously broken in the strongly interacting sector. 
Since our strong sector is supposed to break supersymmetry, a generic 
argument of Ref.~\cite{Nelson:1993nf} suggests that it may possess an 
accidental $U(1)$ $R$ symmetry which is dynamically broken (even if 
supersymmetry is broken in a local minimum).  This then leads to a light 
$R$~axion, whose mass is typically of $O(1~\mbox{--}~100~{\rm MeV})$ 
if the mass dominantly arises from a constant term in the superpotential 
canceling the vacuum energy of order $F^2$~\cite{Bagger:1994hh}.

More generally, it is not hard to imagine that the sector possesses enhanced 
approximate global symmetries.  If these symmetries are spontaneously 
broken, light axion-like states with masses much smaller than the dynamical 
scale will appear.  The masses of these states are then controlled by the 
size of explicit breaking of the corresponding symmetries.
\item{\bf Couplings to the Higgs fields:}
If the supersymmetry breaking sector yields an $R$~axion, it generically 
couples to, and thus mixes with, the Higgs fields.  This is, in fact, the 
case even if the Higgs fields are not directly coupled to the supersymmetry 
breaking sector because the holomorphic Higgs mass-squared ($B\mu$ term) 
obtains loop contributions from gaugino masses, which are necessarily 
$R$-violating.  For other axion-like states, mixing with the Higgs fields 
can arise if the Higgs fields are directly coupled to the supersymmetry 
breaking sector, making the corresponding symmetries Peccei-Quinn (PQ) 
symmetries.  This is well motivated, since such couplings are often 
needed to generate the $\mu$ term.
\end{description}
It is interesting to note that all the ingredients above appear in QCD, 
a known strongly coupled system in nature.  For the first two, one should 
simply think of protons, neutrons, pions, and kaons (in the appropriate 
limits where quark masses are small or weak interactions are superweak). 
Even direct Higgs couplings do exist, although the Higgs boson is much 
heavier than the dynamical scale of QCD, while it is much lighter than 
the dynamical scale considered here.

The structures described above could be manifested in various experiments. 
In particular, couplings between the Higgs and axion-like states provide 
a potential window to probe the supersymmetry breaking sector directly. 
Since the sector may contain dark matter as well as light states, possible 
signatures may appear both in astrophysical and collider physics data, 
and in this paper we consider the following two classes of signatures:
\begin{description}
\item{\bf Cosmic ray signals from decaying dark matter:}
If dark matter is a quasi-stable state in the supersymmetry breaking 
sector, its decay may lead to various astrophysical signatures.  Assuming 
the decay occurs through the light states, the final states can be 
mainly leptons, explaining the excess of the positron to electron 
ratio observed in the PAMELA experiment~\cite{Adriani:2008zr}, along 
the lines of~\cite{Cholis:2008vb,Nomura:2008ru,Chen:2009iua}.  The 
required lifetime of order $10^{26}~{\rm sec}$ is obtained if the decay 
is caused by a dimension six operator suppressed by the unification 
or gravitational scale~\cite{Nardi:2008ix}.

There are characteristic features for this explanation of the PAMELA 
excess which are being tested in current observations.  First, since the 
decay occurs through light states, it typically involves (long) cascade 
chains.  Second, the mass of dark matter is rather large, $m_{\rm DM} 
\approx O(10~{\rm TeV})$, since it arises as a quasi-stable state 
in the supersymmetry breaking sector.  These lead to a rather broad 
structure in the electron plus positron spectrum in the sub-TeV 
region after propagation to the earth.  Remarkably, we find that such 
a structure beautifully reproduces the spectra recently reported by 
the FERMI~\cite{Abdo:2009zk} and H.E.S.S.~\cite{Aharonian:2008aaa,%
Aharonian:2009ah} experiments.
\item{\bf Collider signals of light axion-like states:}
The couplings between the Higgs and axion-like states imply that the 
axion-like states also couple to the standard model gauge fields at the 
loop level.  Such couplings may also arise from contributions from the 
supersymmetry breaking sector.  This leads to the possibility of producing 
the light states at the LHC, which subsequently decay into standard model 
fields through mixings with the Higgs fields~\cite{Goh:2008xz}.  The 
final states are most likely leptons, if recent cosmic ray data are 
explained as described above.  This may provide a relatively clean 
signal to discover the light states.

The couplings of the Higgs to light states also raises the possibility 
that the Higgs boson decays into two axion-like states.  If the axion-like 
state decays mainly into two leptons, then this leads to a four lepton 
final state, whose invariant mass peaks at the Higgs boson mass.  If 
the rate is sufficiently large, this leads to a way of seeing the 
axion-like state (and the Higgs boson) at hadron colliders.
\end{description}
\begin{table}
\begin{center}
\begin{tabular}{|c|c|c|c|}
\hline
  Symmetry & \begin{tabular}{c} Spontaneous \\ Breaking? \end{tabular} 
    & \begin{tabular}{c} Explicit \\ Breaking? \end{tabular} 
    & Consequences \\
\hline \hline
  $R$ or PQ  & Yes & 
    \begin{tabular}{c} Supergravity or \\ $d = 5$ at $M_I$ \end{tabular}& 
    \begin{tabular}{c} Axion-like state with: \\
      $m_a \approx O(1~{\rm MeV}~\mbox{--}~10~{\rm GeV})$ \\
      $f_a \approx O(1~\mbox{--}~100~{\rm TeV})$ \end{tabular} \\
\hline
  $B$ or $F$ &  No & $d = 6$ at $M_*$ & 
    \begin{tabular}{c} Quasi-stable dark matter with: \\
      $m_{\rm DM} \approx O(10~{\rm TeV})$ \\
      $\tau_{\rm DM} \approx O(10^{26}~{\rm sec})$ \end{tabular} \\
\hline
\end{tabular}
\end{center}
\caption{The symmetry structure necessary to realize the scenario 
 presented in this paper.  A light axion-like state emerges from 
 spontaneous breaking of an $R$ or PQ (Peccei-Quinn) symmetry, which 
 then mixes with the Higgs sector of the standard model.  Composite 
 states in the strong sector are quasi-stable because of a $B$ 
 (``baryon number'') or $F$ (``flavor'') symmetry.  The $R$ 
 or PQ symmetry is explicitly broken by supergravity effects 
 or by dimension five operators suppressed by $M_I \approx 
 O(10^9~\mbox{--}~10^{18}~{\rm GeV})$.  This gives a sufficiently 
 large mass to the axion-like state.  Explicit breaking of $B$ or 
 $F$ is due to dimension six operators suppressed by $M_* \approx 
 O(10^{16}~\mbox{--}~10^{18}~{\rm GeV})$, leading to dark matter 
 decay through the light states.}
\label{tab:symmetry}
\end{table}
The symmetry structure necessary to produce the signatures described 
above is summarized in Table~\ref{tab:symmetry}.  While this structure 
is strongly motivated by the low energy supersymmetry breaking scenario, 
all that is actually required is some strong dynamics at $\approx 
O(10~\mbox{--}~100~{\rm TeV})$ satisfying the properties given in 
the table.  Given that QCD already has more or less all the desired 
ingredients, we expect that the required structure may arise naturally 
in wide classes of strongly interacting gauge theories, including 
non-supersymmetric theories.  (For a non-supersymmetric theory, the 
symmetry leading to a light state must be a PQ symmetry.  This implies 
that the theory must have two Higgs doublets.)  The dynamical scale 
of $O(10~\mbox{--}~100~{\rm TeV})$ then suggests that this sector 
is related to the weak scale through a loop factor.  This feature 
is automatic in low energy supersymmetry breaking theories.

A remarkable thing is that the first signature may have already been seen 
in the recent cosmic ray electron/positron data.  The PAMELA experiment 
found an unexpected rise in the positron fraction in the energy range 
between $10$ and $100~{\rm GeV}$, while the FERMI experiment saw an 
excess of the electron plus positron flux over standard diffuse cosmic 
ray backgrounds in the sub-TeV region.  These results suggest a new source 
of primary electrons and positrons with a broad spectrum extending up to 
a few TeV.  We find that these features are very well explained by dark 
matter in our framework: a quasi-stable state with mass of $O(10~{\rm TeV})$, 
cascading into leptons through light axion-like states with lifetime of 
$O(10^{26}~{\rm sec})$.  We perform a detailed analysis for the cosmic 
ray data and find that a wide range for the mass is allowed for the 
axion-like state: it can take any value between $2m_e \simeq 1.0~{\rm MeV}$ 
and $2m_b \simeq 8.4~{\rm GeV}$ except for a small window between 
$2m_p \simeq 1.9~{\rm GeV}$ and $2m_\tau \simeq 3.6~{\rm GeV}$.

These data, therefore, point to the setup of Table~\ref{tab:symmetry} 
as a new paradigm for dark matter, which can be beautifully realized in 
the framework of low energy supersymmetry breaking.  Since the precise 
structure of the supersymmetry breaking sector is highly model dependent, 
one might worry that the signatures considered here may depend on many 
details of the supersymmetry breaking sector, which leads to large 
uncertainties.  This is, however, not the case.  As emphasized above, 
the existence of the signatures depends only on basic symmetry properties 
of the supersymmetry breaking sector, and their characteristics are 
determined only by a few parameters such as the mass of dark matter 
and the mass and decay constant of the axion-like state.  While we 
will provide an illustrative model as a proof-of-concept, many details 
of the model are unimportant for the signatures.  Of course, the 
flip side of this is that we cannot probe the detailed structure 
of the supersymmetry breaking sector solely by studying these 
signatures.  We may, however, still explore some features by 
carefully studying cosmic ray spectra.

The signatures described here are complementary to the information we 
can obtain in other methods.  In the framework of low energy supersymmetry 
breaking, the LHC will be able to measure some of the superparticle masses. 
This, however, may not determine, e.g., the scale of supersymmetry breaking, 
since the most general supersymmetry breaking sector provides little 
definite prediction on superparticle masses, as recently elucidated 
in Ref.~\cite{Meade:2008wd}.  The existence of the very light gravitino can 
give specific signals, for example, those in Ref.~\cite{Dimopoulos:1996vz}. 
The signatures considered here can add even more handles.  In addition 
to indicating the specific symmetry structure of Table~\ref{tab:symmetry}, 
different final states for the axion-like state decay may also be 
discriminated, e.g., by future measurements of the diffuse $\gamma$-ray 
flux at FERMI.  These will provide valuable information in exploring 
the structure of the supersymmetry breaking sector.

The organization of the paper is as follows.  In the next section, we 
describe our supersymmetric setup in detail.  We explain that quasi-stable 
states with the desired lifetimes and light axion-like states with 
the desired masses can naturally arise.  We also discuss constraints 
on axion-like states, and find that a wide range for the masses and 
decay constants are experimentally viable and can lead to leptonic 
decays.  In Section~\ref{sec:illust}, we present an example model 
that illustrates some of these general points.  Dark matter is a stable 
``meson'' state in the hidden sector that decays into $R$~axions with 
a lifetime of $O(10^{26}~{\rm sec})$.  In Section~\ref{sec:astro}, 
we perform a detailed analysis of the recent cosmic ray data, and find 
that the results of PAMELA, FERMI, and H.E.S.S. are very well explained. 
We present a general analysis in the case where dark matter decays 
into $e^+e^-$, $\mu^+\mu^-$, $\pi^+\pi^-\pi^0$, or $\tau^+\tau^-$ 
either directly or through $1$-step or $2$-step cascades.  Implications 
for future diffuse $\gamma$-ray measurements are also discussed.  In 
Section~\ref{sec:collider}, we briefly discuss collider signatures 
associated with the existence of light states.  Finally, discussion 
and conclusions are given in Section~\ref{sec:discuss}, where we 
mention related alternative scenarios.

\section{Framework}
\label{sec:framework}

We consider a supersymmetry breaking sector which consists of fields and 
interactions characterized by a scale
\begin{equation}
  \Lambda \approx O(10~\mbox{--}~100~{\rm TeV}).
\label{eq:Lambda}
\end{equation}
The actual spectrum of this sector could span an order of magnitude or 
so due to its nontrivial structure.  The scale of Eq.~(\ref{eq:Lambda}) 
is supposed to arise dynamically through some strong gauge interactions 
in order to explain why the supersymmetry breaking scale, and 
thus the weak scale, is hierarchically smaller than the Planck 
scale~\cite{Witten:1981nf}.  We assume the sector contains fields 
charged under the standard model gauge group which directly feel 
supersymmetry breaking.  Standard model gauge loops then generate 
gaugino and scalar masses in the supersymmetric standard model 
(SSM) sector through gauge mediation~\cite{Dine:1981gu,Dine:1994vc}.

\begin{figure}[t]
\begin{center}
\begin{picture}(450,105)(0,5)
  \Line(15,20)(15,80)   \CArc(25,80)(10,90,180)
  \Line(25,90)(155,90)  \CArc(155,80)(10,0,90)
  \Line(165,80)(165,20) \CArc(155,20)(10,270,360)
  \Line(25,10)(155,10)  \CArc(25,20)(10,180,270)
  \Text(90,67)[]{Supersymmetric}
  \Text(90,50)[]{standard model}
  \Text(90,33)[]{(SSM) sector}
  \Photon(165,60)(285,60){3}{10}
  \Text(225,87)[b]{Standard model} \Text(225,73)[b]{gauge $g$}
  \DashLine(165,40)(285,40){3}
  \Text(225,31)[t]{Higgs couplings} \Text(225,17)[t]{$\lambda$}
  \Line(285,20)(285,80) \CArc(295,80)(10,90,180)
  \Line(295,90)(425,90) \CArc(425,80)(10,0,90)
  \Line(435,80)(435,20) \CArc(425,20)(10,270,360)
  \Line(295,10)(425,10) \CArc(295,20)(10,180,270)
  \Text(360,67)[]{Supersymmetry}
  \Text(360,50)[]{breaking sector}
  \Text(360,33)[]{$\Lambda \approx (10~\mbox{--}~100)~{\rm TeV}$}
\end{picture}
\caption{A schematic depiction of the setup.}
\label{fig:setup}
\end{center}
\end{figure}
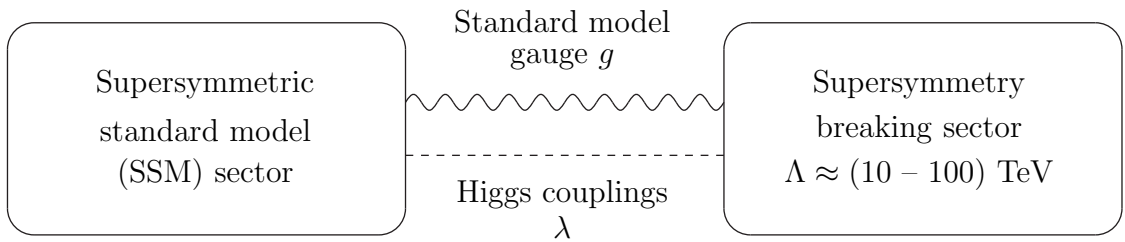

The supersymmetry breaking sector may also directly interact with 
the Higgs fields in the superpotential.  A possible form for these 
interactions is
\begin{equation}
  W = \lambda_u H_u {\cal O}_u + \lambda_d H_d {\cal O}_d,
\label{eq:Higgs-int_1}
\end{equation}
or
\begin{equation}
  W = \lambda N H_u H_d + N {\cal O}_N,
\label{eq:Higgs-int_2}
\end{equation}
where $H_{u,d}$ and $N$ are two Higgs doublet and singlet chiral 
superfields, respectively, and ${\cal O}_{u,d,N}$ represent operators 
in the supersymmetry breaking sector.  These interactions can generate 
the $\mu$ term, and thus lead to realistic electroweak symmetry 
breaking~\cite{Csaki:2008sr,Komargodski:2008ax}.  A schematic picture 
for the current setup can be seen in Figure~\ref{fig:setup}.

We assume that the entire system of Figure~\ref{fig:setup}, including both 
the supersymmetry breaking and SSM sectors (as well as possible direct 
Higgs interactions), respects some approximate global symmetry under which 
{\it both} the supersymmetry breaking and SSM sector fields are charged. 
This implies that the two sectors must have an interaction that can 
transmit charges of the global symmetry from one sector to the other. 
We find that an $R$ symmetry is a natural candidate for such a symmetry, 
since interactions generating gaugino masses must always transmit $R$ 
charges.  Another simple candidate is a PQ symmetry.  In the presence 
of direct Higgs couplings to the supersymmetry breaking sector as in 
Eqs.~(\ref{eq:Higgs-int_1},~\ref{eq:Higgs-int_2}), charges of the PQ 
symmetry can be transmitted.  Note that $R$ and PQ symmetries are global 
symmetries of the SSM sector only in the limit of vanishing $\mu$ term 
and gaugino (and holomorphic supersymmetry breaking) masses.  To generate 
the $\mu$ term and gaugino masses, therefore, these symmetries must be 
spontaneously broken by the dynamics of the supersymmetry breaking sector.

The fundamental supersymmetry breaking scale of this class of theories 
is of order the dynamical scale
\begin{equation}
  \sqrt{F} \approx \Lambda,
\label{eq:sqrt-F_Lambda}
\end{equation}
yielding a light gravitino, $m_{3/2} \approx \Lambda^2/M_{\rm Pl} 
\approx O(0.1~\mbox{--}~10~{\rm eV})$.  This solves all the cosmological 
problems in supersymmetric theories unless the supersymmetry breaking 
sector introduces its own problems,%
\footnote{Possible problems include the system being trapped in the wrong 
 vacuum or the appearance of stable charged or colored particles.}
which we assume not to be the case.  The thermal history of the universe 
is normal up to a very high temperature $T \gg \Lambda$, such that the 
relic abundances of the quasi-stable states are determined by a standard 
thermal freezeout calculation.

\subsection{Quasi-stable states}
\label{subsec:stable}

In addition to the $R$ and PQ symmetries described above, the 
supersymmetry breaking and SSM sectors can have additional independent 
global symmetries.  For example, the SSM sector has a baryon number 
symmetry, at least if $R$~parity is conserved (which we need not assume 
here).  Similarly, the supersymmetry breaking sector may possess accidental 
global symmetries, and since this sector is assumed to feel strong gauge 
interactions, natural possibilities are ``baryon number'' and ``flavor'' 
symmetries.  If the global symmetry is not spontaneously broken by the 
dynamics of this sector, the lightest state charged under that symmetry 
is stable.  This state can then be dark matter, with an abundance 
determined by its annihilation cross section.

Since dark matter is the lightest state charged under a global symmetry, 
it sits at the lowest edge of the spectrum of the corresponding charged 
states.  It is therefore natural to expect that the dark matter mass 
is in the range
\begin{equation}
  m_{\rm DM} \approx O(10~{\rm TeV}).
\label{eq:m_DM}
\end{equation}
In fact, we will see later that a mass of this size reproduces well 
the observed electron/positron spectrum at PAMELA, FERMI, and H.E.S.S. 
experiments.  The annihilation cross section of dark matter is naturally 
of order
\begin{equation}
  \langle \sigma v \rangle 
    \approx \frac{1}{8\pi} \frac{\kappa^4}{m_{\rm DM}^2},
\label{eq:sigma-v}
\end{equation}
where $\kappa$ represents typical couplings between states in the strong 
sector, and we have assumed two-body annihilation.  For $\kappa \approx 3$, 
this gives the cross section needed to reproduce the observed dark matter 
abundance, $\Omega_{\rm DM} \simeq 0.2$.  Such a value for $\kappa$ 
is quite natural for couplings between hadronic states in a strongly 
interacting sector.

The global symmetry ensuring the stability of dark matter is expected 
to be an accidental symmetry at low energies.  This implies that the 
symmetry is not respected by physics at some higher energy $M_*$, 
such as the unification or Planck scale, so that the effective theory 
at the scale $\Lambda$ contains higher dimension operators suppressed by 
powers of $M_*$ that do not respect the global symmetry.  This situation 
is precisely analogous to baryon number in the standard model embedded in 
grand unified theories (with $M_*$ identified with the unification scale), 
or strangeness in QCD (with $M_*$ identified with the weak scale). 
Dark matter can then decay with cosmologically observable timescales, 
depending on the dimension of the leading symmetry-violating operators.

Consistent with gauge coupling unification, we take $M_*$ 
to be around the unification or Planck scale, $M_* \approx 
O(10^{16}~\mbox{--}~10^{18}~{\rm GeV})$.  Then, to have observable 
signatures in cosmic rays, the dimension of relevant symmetry violating 
operators must be six, as illustrated in Ref.~\cite{Nardi:2008ix} in 
the case where $m_{\rm DM} \approx O({\rm TeV})$ and $M_* \approx 
O(10^{16}~{\rm GeV})$.  With dimension six decay operators, we find
\begin{equation}
  \tau_{\rm DM} \approx 8\pi \frac{M_*^4}{m_{\rm DM}^5} 
  \simeq 2 \times 10^{25}~{\rm sec} 
    \left( \frac{M_*}{10^{17}~{\rm GeV}} \right)^4 
    \left( \frac{10~{\rm TeV}}{m_{\rm DM}} \right)^5,
\label{eq:tau-2body}
\end{equation}
for two-body decays, and
\begin{equation}
  \tau_{\rm DM} \approx 128\pi^3 \frac{M_*^4}{m_{\rm DM}^5} 
  \simeq 3 \times 10^{27}~{\rm sec} 
    \left( \frac{M_*}{10^{17}~{\rm GeV}} \right)^4 
    \left( \frac{10~{\rm TeV}}{m_{\rm DM}} \right)^5,
\label{eq:tau-3body}
\end{equation}
for three-body decays.  The required lifetime to fit the PAMELA, FERMI, 
and H.E.S.S. data through dark matter decay is of order $10^{26}~{\rm sec}$, 
which is consistent with the value of $M_*$ taken here.

\subsection{Decay of quasi-stable states}
\label{subsec:decay-stable}

There are several possible ways for the decay of dark matter to be caused 
by dimension six operators.  One simply needs small breaking of the global 
symmetries protecting dark matter, and generically decays will proceed via 
some kinematically-allowed but symmetry-violating channel.  In particular, 
dark matter can decay into light axion-like states and/or the gravitino. 
As long as the dimension six operators are suppressed by the unification 
or Planck scale, they can even explicitly violate the $R$ or PQ symmetry 
that gives rise to the axion-like state, since such explicit breaking 
gives very small masses compared to the contributions considered 
in Section~\ref{subsec:axion}.

Suppose that interactions in the supersymmetry breaking sector are 
asymptotically free, so that the dimensions of the operators are 
determined by the canonical dimensions of the elementary fields. 
Suppose also, for illustrative purposes, that the supersymmetry breaking 
sector contains a gauge group $SU(N_c)$ ($N_c \geq 3$) with $N_f$ flavor 
of ``quark'' fields $Q^i + \bar{Q}_{\bar{\imath}}$ ($i,\bar{\imath} = 
1,\cdots,N_f$), where $Q^i$ and $\bar{Q}_{\bar{\imath}}$ are chiral 
superfields in the fundamental and anti-fundamental representations 
of $SU(N_c)$, respectively.  We also assume that the quark fields 
have generic masses of order $\Lambda$ (or somewhat smaller) due 
to some ($R$-violating) dynamics in the supersymmetry breaking sector.

The $SU(N_c)$ gauge group is supposed to confine at a scale 
$\approx \Lambda$.  This then leads to composite ``meson'' fields 
$M^i_{\bar{\jmath}} \sim Q^i \bar{Q}_{\bar{\jmath}}$, and also ``baryon'' 
and ``antibaryon'' fields $B \sim Q^{N_c}$ and $\bar{B} \sim \bar{Q}^{N_c}$ 
for $N_f \geq N_c$.  Because of the nonzero quark masses, these fields 
have masses of order $\Lambda$ (or somewhat smaller).  Moreover, 
in the limit that nonrenormalizable operators vanish, the states 
$M^i_{\bar{\jmath}}$ ($i \neq \bar{\jmath}$), $B$ and $\bar{B}$ can 
easily be stable (except for $B$ and $\bar{B}$ for $N_c = 3$). 
These states are therefore good candidates for dark matter.

The lifetimes of the quasi-stable states described above are controlled 
by the form of higher dimension operators.  Let us begin with the case 
where dark matter is identified with baryons (and antibaryons).  The 
decay of dark matter can then occur through baryon number violating 
operators in the superpotential.  These operators are dimension six 
if $N_c = 5$:%
\footnote{In the following equations, any one of the operators is 
 sufficient to cause dark matter decay.  In particular, physics at 
 $M_*$ need not respect a $Q \leftrightarrow \bar{Q}$ symmetry.}
\begin{equation}
  W \sim \frac{1}{M_*^2} QQQQQ 
    + \frac{1}{M_*^2} \bar{Q}\bar{Q}\bar{Q}\bar{Q}\bar{Q},
\label{eq:B-vio_1}
\end{equation}
in which case the lifetime of dark matter can be of order $10^{26}~{\rm sec}$ 
as needed to reproduce the electron/positron data.%
\footnote{A long-lived hidden sector baryon with the $SU(5)$ gauge 
 group was considered in Ref.~\cite{Hamaguchi:2008rv} in the context that 
 the quarks $Q^i$ are also charged under the standard model gauge group. 
 This model, however, does not preserve the success of perturbative gauge 
 coupling unification, since the extra matter content charged under the 
 standard model gauge group is too large, see e.g.~\cite{Jones:2008ib}. 
 Here we consider that the quarks $Q^i$ are not charged under the 
 standard model gauge group.}

An alternative possibility is that dark matter decays through holomorphic 
terms in the K\"{a}hler potential $K$, which can also be viewed as 
superpotential terms suppressed by an extra power of $M_{\rm Pl}$ 
after performing a K\"{a}hler transformation.  The required lifetime 
is obtained for $N_c = 3$:
\begin{equation}
  K \sim \frac{1}{M_*} QQQ 
    + \frac{1}{M_*} \bar{Q}\bar{Q}\bar{Q} + {\rm h.c.}
\quad\Longrightarrow\quad
  W \sim \frac{\Lambda^2}{M_* M_{\rm Pl}} QQQ 
    + \frac{\Lambda^2}{M_* M_{\rm Pl}} \bar{Q}\bar{Q}\bar{Q},
\label{eq:B-vio_2}
\end{equation}
where we have used the fact that the superpotential has a constant term 
of order $\Lambda^2 M_{\rm Pl}$ to cancel the cosmological constant, 
and assumed that the renormalizable superpotential term $W \sim QQQ$ 
is absent, perhaps because of an $R$ symmetry.  Note that in both of 
the cases above, the number of ``colors'' $N_c$ needs to be chosen 
appropriately to have dimension six dark matter decay.

We now consider the case where the meson states $M^i_{\bar{\jmath}}$ 
($i \neq \bar{\jmath}$) are dark matter.  These states exist even for 
$N_f < N_c$, and the longevity of their lifetime could be ensured by 
a vector-like $U(1)^{N_f}$ symmetry that may exist in the Lagrangian 
at the renormalizable level.  The decay of these states can be caused 
by dimension six operators in the K\"{a}hler potential:
\begin{equation}
  K \sim \frac{1}{M_*^2} Q^\dagger_i Q^j Q^\dagger_k Q^l 
    + \frac{1}{M_*^2} \bar{Q}^{\dagger\bar{\imath}} 
      \bar{Q}_{\bar{\jmath}} \bar{Q}^{\dagger\bar{k}} \bar{Q}_{\bar{l}} 
    + \frac{1}{M_*^2} Q^\dagger_i Q^j 
      \bar{Q}^{\dagger\bar{k}} \bar{Q}_{\bar{l}}.
\label{eq:F-vio}
\end{equation}
Possible lower dimension operators in the superpotential $W \sim 
(1/M_*) Q^i \bar{Q}_{\bar{\jmath}} Q^k \bar{Q}_{\bar{l}}$ can easily 
be absent, for example by imposing an $R$ symmetry.  An attractive 
feature of this possibility is that the lifetime does not depend on 
the number of colors $N_c$, and that the number of flavors $N_f$ need 
not be equal to or larger than $N_c$.  This setup, therefore, can 
naturally be accommodated in a wide variety of gauge theories. 
In Section~\ref{sec:illust}, we present an explicit model realizing 
this possibility, where the quasi-stable mesons decay via 
Eq.~(\ref{eq:F-vio}) into $R$~axions.

Although we have not considered them in the discussion above, in general 
the supersymmetry breaking sector also contains states not charged 
under $SU(N_c)$.  Dark matter decay may then occur through these states. 
For example, mesons may decay through operators of the form $K \sim 
Q^\dagger_i Q^j \Phi^\dagger \Phi/M_*^2 + \bar{Q}^{\dagger\bar{\imath}} 
\bar{Q}_{\bar{\jmath}} \Phi^\dagger \Phi/M_*^2$, where $\Phi$ represents 
a generic field in the supersymmetry breaking sector.  The existence 
of singlet fields can also change the requirement on $N_c$ for the 
baryon dark matter case.  As long as the lowest symmetry breaking 
operators are dimension six, however, the existence of these operators 
do not affect the basic argument.%
\footnote{With singlets, some of the accidental symmetries are not 
 as ``automatic'' as the case without singlets.  For example, the 
 low energy $U(1)^{N_f}$ flavor symmetry does not exist if the quarks 
 couple to two or more singlets with arbitrary Yukawa couplings.}
Below, we assume for simplicity that operators containing only quarks 
dominate the decay.  Whether this is the case or not is determined by 
physics at the scale $M_*$.  This also implies that direct interactions 
between the supersymmetry breaking and SSM sectors, such as $K \sim 
Q^\dagger_i Q^j \Phi_{\rm SSM}^\dagger \Phi_{\rm SSM}/M_*^2 + 
\bar{Q}^{\dagger\bar{\imath}} \bar{Q}_{\bar{\jmath}} \Phi_{\rm SSM}^\dagger 
\Phi_{\rm SSM}/M_*^2$, are relatively suppressed.  Here, $\Phi_{\rm SSM}$ 
represents SSM fields.  We will comment on the case where these 
interactions are relevant in Section~\ref{sec:discuss}.

\subsection{Light axion-like states}
\label{subsec:axion}

Spontaneous breaking of an approximate $R$ or PQ symmetry in the 
supersymmetry breaking sector leads to a pseudo Nambu-Goldstone boson. 
The phenomenology associated with this particle is determined largely 
by its mass $m_a$ and decay constant $f_a$.  Suppose that relevant 
interactions in the supersymmetry breaking sector obey naive dimensional 
analysis for a generic strongly coupled theory~\cite{Luty:1997fk}. 
In this case, we find that the decay constant is naturally somewhat 
($\approx 4\pi$) smaller than the dynamical scale $\Lambda \approx 
O(10~\mbox{--}~100~{\rm TeV})$:
\begin{equation}
  f_a \approx O(1~\mbox{--}~10~{\rm TeV}).
\label{eq:f_a}
\end{equation}
According to naive dimensional analysis, the generic size for expectation 
values of the lowest and highest components of a chiral superfield $\Phi$ 
are given by $\langle \Phi \rangle \approx \Lambda/4\pi$ and $\langle 
F_\Phi \rangle \approx \Lambda^2/4\pi$, respectively, and a generic 
coupling constant has the size $\kappa \approx 4\pi$.  (Here we have 
ignored a possible $N_c$ or $N_f$ factor associated with the multiplicity 
of fields, but this does not affect the basic discussion.)  This implies 
that a generic supersymmetric mass and supersymmetry breaking mass-squared 
splitting in the supersymmetry breaking sector are of order $M_{\rm mess} 
\approx \kappa \langle \Phi \rangle \approx \Lambda$ and $F_{\rm mess} 
\approx \kappa \langle F_\Phi \rangle \approx \Lambda^2$, respectively, 
so that the gaugino and scalar masses generated in the SSM sector are 
of order $(g^2/16\pi^2) F_{\rm mess}/M_{\rm mess} \approx (g^2/16\pi^2) 
\Lambda$, which is consistent with Eq.~(\ref{eq:Lambda}).  On the other 
hand, the decay constant of an axion-like state scales as $f_a \approx 
\langle \Phi \rangle \approx \Lambda/4\pi$, giving Eq.~(\ref{eq:f_a}). 
Note that the suppression of $f_a$ over $\Lambda$ here is precisely 
analogous to the fact that in QCD the pion decay constant, $f_\pi \approx 
O(100~{\rm MeV})$, is an order of magnitude smaller than the characteristic 
QCD scale, i.e.\ the rho meson mass $m_\rho \approx O(1~{\rm GeV})$.%
\footnote{In general, it may not be true that all the couplings in the 
 supersymmetry breaking sector are strong.  In this case, the decay 
 constant need not obey Eq.~(\ref{eq:f_a}); for example, it can be 
 easily of $O(100~{\rm TeV})$.  In particular, if the messenger fields 
 (fields in the supersymmetry breaking sector charged under the standard 
 model gauge group) feel $R$-breaking effects though perturbative 
 interactions, then $f_a$ for an $R$~axion needs to be of $O(10~{\rm TeV})$ 
 or larger to generate sufficiently large gaugino masses.}

The mass of a light axion-like state is determined by the size of explicit 
symmetry breaking.  Let us first consider the state associated with an 
$R$ symmetry---an $R$~axion.  An interesting feature of an $R$~axion is 
that it has an irreducible contribution to its mass from supergravity, 
whose size can be determined by the strength of fundamental supersymmetry 
breaking $F$.  This is because to cancel the cosmological constant, the 
superpotential must have a constant piece $\langle W \rangle \approx 
F M_{\rm Pl} \approx O(\Lambda^2 M_{\rm Pl})$, which is necessarily 
$R$-violating.  Since $\langle W \rangle \gg \Lambda^3$, the constant 
piece must come from a sector other than the supersymmetry breaking 
sector, implying that it appears as explicit breaking from the perspective 
of the supersymmetry breaking sector.  This provides the following 
contribution to the $R$~axion mass~\cite{Bagger:1994hh}
\begin{equation}
  m_a^2 \approx 4\pi \frac{F \langle \Phi \rangle^3}{M_{\rm Pl} f_a^2} 
    \approx \frac{\Lambda^3}{4\pi M_{\rm Pl}} 
    \approx O(10~{\rm MeV})^2 \left( \frac{\Lambda}{100~{\rm TeV}} \right)^3,
\label{eq:m_a-R}
\end{equation}
where $\langle \Phi \rangle$ is a generic vacuum expectation value 
in the supersymmetry breaking sector, and we have used $\langle \Phi 
\rangle \approx \Lambda/4\pi$, $f_a \approx \Lambda/4\pi$, and 
$F \approx \Lambda^2/4\pi$.  The uncertainty of the estimate, however, 
is very large, so that we can easily imagine $m_a$ in the range 
$O(1~\mbox{--}~100~{\rm MeV})$.  Unless the supersymmetry breaking 
sector contains another explicit breaking of the $R$ symmetry, the 
$R$~axion mass is given by Eq.~(\ref{eq:m_a-R}).

We now consider the case of a PQ symmetry, or of additional explicit 
breaking of an $R$ symmetry.  If the symmetry is violated by dimension 
five operators in the supersymmetry breaking sector, then we expect
\begin{equation}
  m_a^2 \approx \frac{f_a^3}{M_I} 
    \approx O(100~{\rm MeV})^2 
      \left( \frac{\Lambda}{100~{\rm TeV}} \right)^3 
      \left( \frac{10^{14}~{\rm GeV}}{M_I} \right),
\label{eq:m_a-PQ}
\end{equation}
where we have introduced the scale $M_I$ for physics causing the explicit 
breaking.  We imagine $M_I$ to take a value between an intermediate scale 
and the gravitational scale, $M_I \approx O(10^9~\mbox{--}~10^{18}~{\rm 
GeV})$, giving $m_a \approx O(1~{\rm MeV}~\mbox{--}~10~{\rm GeV})$. 
While smaller masses are possible, they are in conflict with astrophysical 
measurements as we will see later.  The origin of the intermediate scale 
might be associated, for example, with the Peccei-Quinn scale for the QCD 
axion, the scale of the constant term in the superpotential, or the $B-L$ 
breaking scale generating right-handed neutrino masses.%
\footnote{If we use $f_a \approx \Lambda$ instead of $\Lambda/4\pi$, 
 we obtain $m_a^2 \approx O(100~{\rm MeV})^2 (\Lambda/100~{\rm TeV})^3 
 (10^{17}~{\rm GeV}/M_I)$.  This gives $m_a \approx 
 O(1~{\rm MeV}~\mbox{--}~1~{\rm GeV})$ for $M_I \approx M_*$.}
Alternatively, the explicit breaking may be due to a tiny dimensionless 
coupling in the supersymmetry breaking sector.  Note that a possible 
contribution from the QCD anomaly, $m_a \approx m_\pi (f_\pi/f_a) 
\approx O(1~{\rm keV}) (100~{\rm TeV}/\Lambda)$, is small.

\subsection{Couplings of axion-like states}
\label{subsec:axi-coupling}

The couplings of the axion-like states to the SSM sector are completely 
determined by the symmetry structure.  We are considering the case in which 
the Higgs bilinear $h_u h_d$ is charged under the symmetry that leads to 
the light state, where $h_{u,d}$ are the lowest components of $H_{u,d}$. 
This is almost always true for an $R$ symmetry (unless direct couplings 
of the Higgs to the supersymmetry breaking sector force vanishing charges 
for $H_{u,d}$), and is by definition true for a PQ symmetry.  In both cases, 
we can take a field basis in which the axion-like state $a$ is mixed into 
the $h_{u,d}$ fields:
\begin{eqnarray}
  h_u &=& v_u\, e^{i\frac{\cos^2\!\!\beta}{\sqrt{2}f_a} a},
\label{eq:axion-h_u}\\
  h_d &=& v_d\, e^{i\frac{\sin^2\!\!\beta}{\sqrt{2}f_a} a},
\label{eq:axion-h_d}
\end{eqnarray}
where $v_{u,d} = \langle h_{u,d} \rangle$ and $\tan\beta = \langle h_u 
\rangle/\langle h_d \rangle$.  The distribution of $a$ inside $h_{u,d}$ 
is determined by the condition that $a$ is orthogonal to the mode 
absorbed by the $Z$ boson, and we have arbitrary chosen the $O(1)$ 
normalization for $f_a$.  The expressions of Eqs.~(\ref{eq:axion-h_u},%
~\ref{eq:axion-h_d}) completely determine the leading-order couplings 
of $a$ to the standard model quarks and leptons.  For example, the 
couplings to the up quark, down quark, and electron relevant for $a$ 
decay are given by
\begin{equation}
  {\cal L} = - i\frac{m_u \cos^2\!\beta}{\sqrt{2} f_a}\, 
      a\, \bar{\Psi}_u \gamma_5 \Psi_u 
    - i\frac{m_d \sin^2\!\beta}{\sqrt{2} f_a}\, 
      a\, \bar{\Psi}_d \gamma_5 \Psi_d 
    - i\frac{m_e \sin^2\!\beta}{\sqrt{2} f_a}\, 
      a\, \bar{\Psi}_e \gamma_5 \Psi_e,
\label{eq:a-coupling-f}
\end{equation}
where $m_{u,d,e}$ are the up quark, down quark, and electron masses. 
The couplings to heavier generation fermions are similar.

The couplings of $a$ to the Higgs boson $h$ can be obtained by replacing 
$v_{u,d}$ as
\begin{equation}
  v_u \rightarrow v_u + \frac{\cos\alpha}{\sqrt{2}} h,
\qquad
  v_d \rightarrow v_d - \frac{\sin\alpha}{\sqrt{2}} h,
\label{eq:higgs}
\end{equation}
in Eqs.~(\ref{eq:axion-h_u},~\ref{eq:axion-h_d}) and plugging the resulting 
expressions into the Higgs kinetic terms ${\cal L} = |\partial_\mu h_u|^2 
+ |\partial_\mu h_d|^2$.  Here, $\alpha$ is the Higgs mixing angle.  This 
leads to
\begin{equation}
  {\cal L} = \frac{c_1 v}{\sqrt{2} f_a^2}\, h (\partial_\mu a)^2 
    + \frac{c_2}{4 f_a^2}\, h^2 (\partial_\mu a)^2,
\label{eq:a-coupling-h}
\end{equation}
where $v = \sqrt{v_u^2+v_d^2}$, $c_1 = \sin\beta \cos\beta (\cos^3\!\beta 
\cos\alpha - \sin^3\!\beta \sin\alpha)$, and $c_2 = \cos^4\!\beta 
\cos^2\!\alpha + \sin^4\!\beta \sin^2\!\alpha$.  In the decoupling 
limit, $\alpha \approx \beta - \pi/2$, this equation reduces to
\begin{equation}
  {\cal L} = \frac{v \sin^2\!2\beta}{4\sqrt{2} f_a^2}\, h (\partial_\mu a)^2 
    + \frac{\sin^2\!2\beta}{16 f_a^2}\, h^2 (\partial_\mu a)^2.
\label{eq:a-coupling-h_dec}
\end{equation}
The first term is responsible for Higgs decay into two axion-like states.

There are also couplings between $a$ and the standard model gauge bosons. 
Their precise values depend on the matter content of the entire theory. 
At a given energy scale $E$, effective direct interactions between $a$ 
and the gauge bosons are given by
\begin{equation}
  {\cal L} = \sum_A \frac{g_A^2 c_A}{32\pi^2 \sqrt{2} f_a}\, 
    a F^A_{\mu\nu} \tilde{F}^{A \mu\nu},
\label{eq:a-coupling-gauge}
\end{equation}
where $A$ runs over the gauge groups accessible at the scale $E$, and 
$g_A$ and $c_A$ are the gauge coupling and a coefficient of order unity. 
The coefficient $c_A$ encodes the contributions from physics above $E$, 
and is determined by the symmetry properties of the fields integrated 
out to obtain the effective theory.  At the electroweak scale or below, 
we generically expect $c_A \neq 0$ for color, $A = SU(3)_C$, and 
electromagnetism, $A = U(1)_{\rm EM}$.  The interaction with gluons 
($A = SU(3)_C$) is responsible for direct production of $a$ at hadron 
colliders.

\subsection{Constraints on axion-like states}
\label{subsec:axi-const}

The axion-like states for the $R$ and PQ symmetries considered here 
couple to the standard model fields as the DFSZ axion~\cite{Dine:1981rt}, 
except for possible differences in the numerical coefficients $c_A$. 
Their masses, however, are much heavier because of explicit symmetry 
breaking, so that the experimental constraints on them are quite 
different from those on the DFSZ axion.  Here we summarize the constraints 
on $m_a$ and $f_a$ for the region relevant to our discussions.  For 
previous related analyses, see Refs.~\cite{Nomura:2008ru,Hall:2004qd}.

The constraints on an axion-like state $a$ depend strongly on its 
decay mode.  The decay width of $a$ into two fermions is given by
\begin{equation}
  \Gamma(a \rightarrow f\bar{f}) 
  = \frac{n_f c_f^2}{16\pi} \frac{m_f^2 m_a}{f_a^2} 
    \left( 1 - \frac{4 m_f^2}{m_a^2} \right)^{1/2},
\label{eq:a-ff}
\end{equation}
where $n_f = 1$ and $3$ for leptons and quarks, respectively, and $c_f 
= \sin^2\!\beta$ for $f=e,\mu,\tau,d,s,b$ while $c_f = \cos^2\!\beta$ 
for $f=u,c,t$.  The decay width into two photons is
\begin{equation}
  \Gamma(a \rightarrow \gamma\gamma) 
  = \frac{c_\gamma^2 e^4}{32\pi(16\pi^2)^2} \frac{m_a^3}{f_a^2},
\label{eq:a-gg}
\end{equation}
where $c_\gamma$ represents the $c_A$ coefficient for $U(1)_{\rm EM}$ 
at energies below $m_e$, and $e$ is the electromagnetic gauge coupling.

For $m_a < 2 m_e$, $a$ decays mainly into two photons with the decay 
width of Eq.~(\ref{eq:a-gg}), giving
\begin{equation}
  c\tau_{a \rightarrow \gamma\gamma} \simeq 5.9 \times 10^9~{\rm m}\,\, 
    \frac{1}{c_\gamma^2} \left( \frac{1~{\rm MeV}}{m_a} \right)^3 
    \left( \frac{f_a}{10~{\rm TeV}} \right)^2.
\label{eq:ct-gg}
\end{equation}
A possible decay of an $R$~axion into two gravitinos has the width 
$\Gamma(a \rightarrow \tilde{G}\tilde{G}) \approx (1/16\pi) m_{3/2}^2 
m_a/f_a^2$, and is thus negligible unless $m_a \simlt (4\pi/\alpha) 
m_{3/2}$.  For $2m_e \simlt m_a \simlt 2m_\mu$, $a$ decays dominantly 
into $e^+ e^-$ with the decay width of Eq.~(\ref{eq:a-ff}), giving
\begin{equation}
  c\tau_{a \rightarrow e^+e^-} \simeq 3.8 \times 10^2~{\rm m}\,\, 
    \frac{1}{\sin^4\!\beta} \left( \frac{10~{\rm MeV}}{m_a} \right) 
    \left( \frac{f_a}{10~{\rm TeV}} \right)^2 
    \left( 1 - \frac{4 m_e^2}{m_a^2} \right)^{-1/2}.
\label{eq:ct-ee}
\end{equation}
For $2m_\mu < a \simlt 800~{\rm MeV}$, the $a \rightarrow \mu^+\mu^-$ 
mode dominates with
\begin{equation}
  c\tau_{a \rightarrow \mu^+\mu^-} \simeq 3.0 \times 10^{-4}~{\rm m}\,\, 
    \frac{1}{\sin^4\!\beta} \left( \frac{300~{\rm MeV}}{m_a} \right) 
    \left( \frac{f_a}{10~{\rm TeV}} \right)^2 
    \left( 1 - \frac{4 m_\mu^2}{m_a^2} \right)^{-1/2}.
\label{eq:ct-mumu}
\end{equation}
In this region, the $a \rightarrow \pi\pi$ mode is suppressed by 
$CP$ invariance, and $a \rightarrow \pi\pi\pi$ has the width of order 
$(1/128\pi^3)(m_\pi^4 m_a/f_\pi^2 f_a^2)$, which is much smaller than 
$\Gamma(a \rightarrow \mu^+\mu^-)$.  For $m_a \simgt 800~{\rm MeV}$, 
the $a \rightarrow \rho^*\pi \rightarrow \pi\pi\pi$ and $a \rightarrow 
\eta\pi\pi$ modes become important, but the final states still contain 
a significant fraction of leptons from charged pion decay, unless 
$m_a \simgt 2~{\rm GeV}$ where nucleon modes start dominating. 
For $2m_\tau < m_a < 2m_b$, $a$ will decay dominantly into taus with
\begin{equation}
  c\tau_{a \rightarrow \tau^+\tau^-} \simeq 6.3 \times 10^{-8}~{\rm m}\,\, 
    \frac{1}{\sin^4\!\beta} \left( \frac{5~{\rm GeV}}{m_a} \right) 
    \left( \frac{f_a}{10~{\rm TeV}} \right)^2 
    \left( 1 - \frac{4 m_\tau^2}{m_a^2} \right)^{-1/2}.
\label{eq:ct-tautau}
\end{equation}
The branching ratio into $c\bar{c}$ is $\approx 3 m_c^2/m_\tau^2 
\tan^4\!\beta$, which is highly suppressed for $\tan\beta \simgt 2$. 
For $m_a > 2m_b$ the $a \rightarrow b\bar{b}$ mode dominates.

Rare decays of mesons provide strong constraints on axion-like states. 
In particular, the $K^+ \rightarrow \pi^+ a$ process gives significant 
constraints on the region of interest.  The theoretical estimate 
for the branching ratio is ${\rm Br}(K^+ \rightarrow \pi^+ a) \simgt 
1.1 \times 10^{-8} (10~{\rm TeV}/f_a)^2$~\cite{Antoniadis:1981zw}.  For 
$m_a < 2m_\mu$, the $a$ decay length is large enough that $a$ appears 
as an invisible particle in $K$ decay experiments.  The experimental 
bound on the branching ratio is then ${\rm Br}(K^+ \rightarrow \pi^+ a) 
\simlt 7.3 \times 10^{-11}$~\cite{Anisimovsky:2004hr}.  While the 
theoretical estimate has large uncertainties, this gives the rough 
bound
\begin{equation}
  f_a \simgt 100~{\rm TeV}
\quad\mbox{for}\quad
  m_a < 2 m_\mu.
\label{eq:a-const-RD-1}
\end{equation}
For $2m_\mu < m_a < m_K - m_\pi$, $a$ decays quickly into $\mu^+ \mu^-$, 
so that the relevant experimental data is ${\rm Br}(K^+ \rightarrow 
\pi^+ \mu^+ \mu^-) \simeq 1 \times 10^{-7}$~\cite{Park:2001cv}, which 
is consistent with standard model expectations.  Considering that 
the dimuon invariant mass would be peaked at $m_a$ for $a$ decay, 
${\rm Br}(K^+ \rightarrow \pi^+ a)$ should be somewhat smaller than 
this number, giving a conservative bound of
\begin{equation}
  f_a \simgt \mbox{a few}~{\rm TeV}
\quad\mbox{for}\quad
  2 m_\mu < m_a < m_K - m_\pi.
\label{eq:a-const-RD-2}
\end{equation}
Radiative decay of $\Upsilon$ also provides constraints, but the 
bounds are typically $f_a \simgt O({\rm TeV})$ (for $a \rightarrow 
\mu^+\mu^-$~\cite{Aubert:2009cp}) or weaker (for $a \rightarrow 
\tau^+\tau^-$~\cite{Love:2008hs}), and do not significantly constrain 
the parameter region considered here.

There are constraints from beam-dump experiments.  For example, 
the experiment of Ref.~\cite{Bergsma:1985qz} excludes $f_a \simlt 
(10~\mbox{--}~100)~{\rm TeV}$ for $2 m_e < m_a < 2 m_\mu$.  None of 
these experiments, however, gives as strong bounds as the ones from 
kaon decay given above, except for possible small islands in parameter 
space.  Constraints from reactor experiments are also similar.  They 
are not as strong as those from kaon decay, except that the experiment 
of Ref.~\cite{Altmann:1995bw} excludes a region of $f_a$ somewhat above 
$100~{\rm TeV}$ for $2 m_e < m_a \simlt 10~{\rm MeV}$.

Astrophysics provides strong constraints on very light axion-like 
states~\cite{Raffelt:1990yz}.  A combination of the bounds from the 
dynamics of the sun, white dwarfs, and horizontal branch stars excludes 
$m_a \simlt 300~{\rm keV}$ for the relevant region of $f_a \approx 
O(1~\mbox{--}~100~{\rm TeV})$.  Supernova 1987A could also provide 
potentially strong constraints.  If an $a$ produced in a supernovae 
were to freely escape, then its mass would be excluded up to $\approx 
O(1~{\rm GeV})$ with the $a$ production rates corresponding to the 
$f_a$ values considered here.%
\footnote{We thank Savas Dimopoulos for discussion on this point.}
This is, however, not the case.  For $f_a \approx 
O(1~\mbox{--}~100~{\rm TeV})$, the produced $a$ is either trapped 
inside the supernovae or decays quickly, so that it does not carry 
away significant energy.  In particular, for $m_a > 2 m_\mu$, $a$ 
immediately decays into muons, which are then thermalized quickly. 
Supernova 1987A, therefore, does not strongly constrain the parameter 
space considered here, except that the region $m_a < 2 m_e$ 
is excluded by nonobservation of $\gamma$ rays from $a$ 
decays~\cite{Engel:1990zd}.

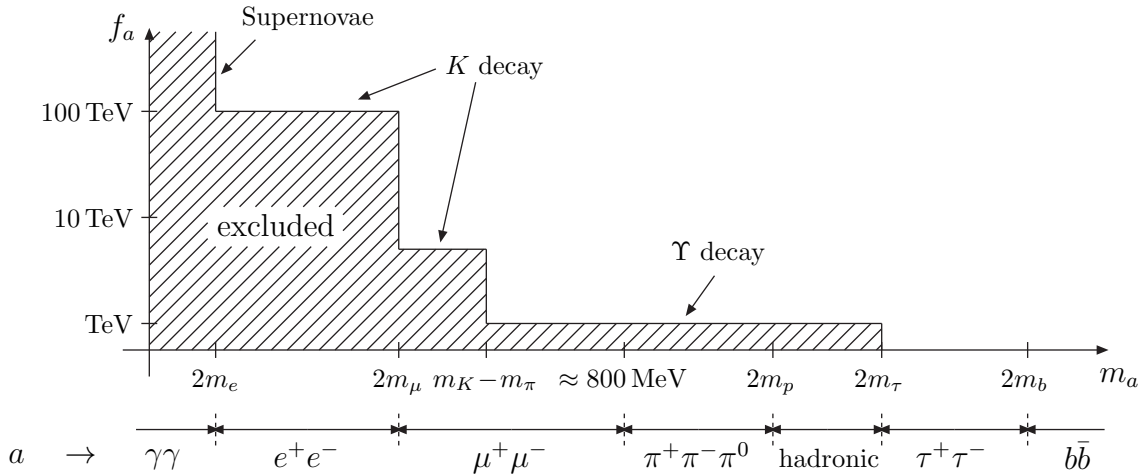
\begin{figure}[t]
\begin{center}
\begin{picture}(370,195)(-25,-50)
  \LongArrow(-10,0)(360,0) \Text(367,-7)[t]{$m_a$}
  \Line(25,-3)(25,3) \Text(25,-7.6)[t]{\footnotesize $2 m_e$}
  \Line(94,-3)(94,3) \Text(94,-7.6)[t]{\footnotesize $2 m_\mu$}
  \Line(127,-3)(127,3) \Text(127,-8)[t]{\footnotesize $m_K\!-\!m_\pi$}
  \Line(179,-3)(179,3) \Text(179,-7)[t]{\footnotesize $\approx 800\,{\rm MeV}$}
  \Line(235,-3)(235,3) \Text(235,-7.6)[t]{\footnotesize $2 m_p$}
  \Line(276,-3)(276,3) \Text(276,-7.6)[t]{\footnotesize $2 m_\tau$}
  \Line(331,-3)(331,3) \Text(331,-7.6)[t]{\footnotesize $2 m_b$}
  \LongArrow(0,-10)(0,120) \Text(-5,122)[r]{$f_a$}
  \Line(-3,10)(3,10) \Text(-6,10)[r]{\footnotesize ${\rm TeV}$}
  \Line(-3,50)(3,50) \Text(-6,50)[r]{\footnotesize $10\,{\rm TeV}$}
  \Line(-3,90)(3,90) \Text(-6,90)[r]{\footnotesize $100\,{\rm TeV}$}
  \Line(25,120)(25,90)
  \Line(25,90)(94,90)
  \Line(94,90)(94,38)
  \Line(94,38)(127,38)
  \Line(127,38)(127,10)
  \Line(127,10)(276,10)
  \Line(276,10)(276,0)
  \LongArrow(40,117)(28,105) \Text(35,121)[lb]{\footnotesize Supernovae}
  \LongArrow(108,106)(80,95) \Text(112,107)[l]{\footnotesize $K$ decay}
  \LongArrow(123,101)(110,42)
  \LongArrow(213,28)(203,14) \Text(215,32)[b]{\footnotesize $\Upsilon$ decay}
  \Line(4,120)(0,116)
  \Line(10,120)(0,110)
  \Line(16,120)(0,104)
  \Line(22,120)(0,98)
  \Line(25,117)(0,92)
  \Line(25,111)(0,86)
  \Line(25,105)(0,80)
  \Line(25,99)(0,74)
  \Line(25,93)(0,68)
  \Line(28,90)(0,62)
  \Line(34,90)(0,56)
  \Line(40,90)(0,50)
  \Line(46,90)(0,44)
  \Line(52,90)(0,38)
  \Line(58,90)(0,32)
  \Text(48,47.2)[]{excluded}
  \Line(64,90)(28,54) \Line(22,48)(0,26)
  \Line(70,90)(34,54) \Line(22,42)(0,20)
  \Line(76,90)(40,54) \Line(25,39)(0,14)
  \Line(82,90)(46,54) \Line(31,39)(0,8)
  \Line(88,90)(52,54) \Line(37,39)(0,2)
  \Line(94,90)(58,54) \Line(43,39)(4,0)
  \Line(94,84)(64,54) \Line(49,39)(10,0)
  \Line(94,78)(70,54) \Line(55,39)(16,0)
  \Line(94,72)(73,51) \Line(61,39)(22,0)
  \Line(94,66)(73,45) \Line(67,39)(28,0)
  \Line(94,60)(34,0)
  \Line(94,54)(40,0)
  \Line(94,48)(46,0)
  \Line(94,42)(52,0)
  \Line(96,38)(58,0)
  \Line(102,38)(64,0)
  \Line(108,38)(70,0)
  \Line(114,38)(76,0)
  \Line(120,38)(82,0)
  \Line(126,38)(88,0)
  \Line(127,33)(94,0)
  \Line(127,27)(100,0)
  \Line(127,21)(106,0)
  \Line(127,15)(112,0)
  \Line(128,10)(118,0)
  \Line(134,10)(124,0)
  \Line(140,10)(130,0)
  \Line(146,10)(136,0)
  \Line(152,10)(142,0)
  \Line(158,10)(148,0)
  \Line(164,10)(154,0)
  \Line(170,10)(160,0)
  \Line(176,10)(166,0)
  \Line(182,10)(172,0)
  \Line(188,10)(178,0)
  \Line(194,10)(184,0)
  \Line(200,10)(190,0)
  \Line(206,10)(196,0)
  \Line(212,10)(202,0)
  \Line(218,10)(208,0)
  \Line(224,10)(214,0)
  \Line(230,10)(220,0)
  \Line(236,10)(226,0)
  \Line(242,10)(232,0)
  \Line(248,10)(238,0)
  \Line(254,10)(244,0)
  \Line(260,10)(250,0)
  \Line(266,10)(256,0)
  \Line(272,10)(262,0)
  \Line(276,8)(268,0)
  \Line(276,2)(274,0)
  \DashLine(25,-24)(25,-36){2}
  \DashLine(94,-24)(94,-36){2}
  \DashLine(179,-24)(179,-36){2}
  \DashLine(235,-24)(235,-36){2}
  \DashLine(276,-24)(276,-36){2}
  \DashLine(331,-24)(331,-36){2}
  \Text(-20,-37.5)[tr]{$a\quad\rightarrow$}
  \LongArrow(-5,-30)(25,-30)
  \Text(5,-37.5)[t]{$\gamma\gamma$}
  \LongArrow(59.5,-30)(25,-30) \LongArrow(59.5,-30)(94,-30)
  \Text(59.5,-35)[t]{$e^+ e^-$}
  \LongArrow(136.5,-30)(94,-30) \LongArrow(136.5,-30)(179,-30)
  \Text(136.5,-35)[t]{$\mu^+ \mu^-$}
  \LongArrow(207,-30)(179,-30) \LongArrow(207,-30)(235,-30)
  \Text(207,-35)[t]{$\pi^+ \pi^- \pi^0$}
  \LongArrow(255.5,-30)(235,-30) \LongArrow(255.5,-30)(276,-30)
  \Text(257,-37.5)[t]{\footnotesize hadronic}
  \LongArrow(303.5,-30)(276,-30) \LongArrow(303.5,-30)(331,-30)
  \Text(303.5,-35)[t]{$\tau^+ \tau^-$}
  \LongArrow(360,-30)(331,-30)
  \Text(351,-35)[t]{$b\bar{b}$}
\end{picture}
\caption{A schematic picture for the constraints on the $m_a$-$f_a$ plane, 
 with the shaded region corresponding to the excluded region.  Note that 
 the actual limits on $f_a$ have $O(1)$ uncertainties, as explained in 
 the text.  The dominant $a$ decay mode for a given value of $m_a$ is 
 also depicted.}
\label{fig:ma-fa}
\end{center}
\end{figure}
A schematic diagram summarizing the bounds on $m_a$ and $f_a$ is 
depicted in Figure~\ref{fig:ma-fa}, together with the dominant $a$ 
decay modes.  For $f_a \approx O(1~\mbox{--}~100~{\rm TeV})$, the most 
natural region avoiding all the constraints are $m_a > 2 m_\mu$, although 
$m_a < 2 m_e$ is still allowed for $f_a \approx O(100~{\rm TeV})$. 
For the range $m_a \approx O(1~{\rm MeV}~\mbox{--}~10~{\rm GeV})$ 
considered in Eqs.~(\ref{eq:m_a-R},~\ref{eq:m_a-PQ}), a natural $a$ 
decay mode will be either $\mu^+\mu^-$, $\tau^+\tau^-$, $\pi^+\pi^-\pi^0$, 
or hadronic (or $e^+e^-$ for $f_a \approx O(100~{\rm TeV})$).  As we 
will see in Section~\ref{sec:astro}, $a$ decay into $\mu^+\mu^-$, 
$\tau^+\tau^-$, or $\pi^+\pi^-\pi^0$ (or $e^+e^-$) can provide 
a spectacular explanation for the recently observed electron/positron 
signals in cosmic rays.

\section{Illustration}
\label{sec:illust}

In this section, we present an example model which illustrates some 
of the general points discussed above.  The model has quasi-stable 
meson fields carrying nontrivial flavor, which we identify with dark 
matter.  We find that this dark matter, in fact, decays dominantly 
into $R$~axions (an $R$~axion and a gravitino if the lightest meson 
field is a fermion) through dimension six operators.  The lifetime 
of dark matter can therefore be naturally of $O(10^{26}~{\rm sec})$. 
The resulting $R$~axion decays into standard model particles as 
shown in Section~\ref{subsec:axi-coupling}, leading to astrophysical 
signatures discussed below in Section~\ref{sec:astro}.  We also show 
that under reasonable assumptions, the correct thermal abundance for 
dark matter can be obtained through a standard freezeout calculation.

\subsection{Setup in supersymmetric QCD}
\label{subsec:setup}

Ideally, one would want to construct a complete model where supersymmetry 
(and $R$ symmetry) is dynamically broken and successfully mediated 
to the SSM sector.  It is, however, notoriously difficult to realize 
such explicit constructions~\cite{Giudice:1998bp}.  Here, we will use 
the example of supersymmetric QCD to show that quasi-stable dark matter 
candidates can arise from strong dynamics with spontaneous $R$-symmetry 
breaking.  In Section~\ref{subsec:more_comp}, we will show that this 
strong sector can be consistently coupled to supersymmetry breaking 
dynamics and to messenger fields, and we will comment on the parametric 
difference between the dark matter mass and the scale of supersymmetry 
breaking.

Our starting point here is supersymmetric QCD where the number 
of flavors $N_f$ is less than the number of colors $N_c$.  In the 
ultraviolet, the matter content consists of quark fields $Q^i$ 
and $\bar{Q}_{\bar{\imath}}$ ($i,\bar{\imath} = 1,\cdots,N_f$). 
We couple these fields to a singlet field $S$ in the superpotential:
\begin{equation}
  W = \lambda_i^{\bar{\jmath}} S Q^i \bar{Q}_{\bar{\jmath}}.
\label{eq:SQQ}
\end{equation}
With only one singlet, these couplings can always be diagonalized by 
rotations of the quark fields, $\lambda_i^{\bar{\jmath}} = \lambda_i 
\delta_i^{\bar{\jmath}}$, and we can then easily see that this theory 
has an accidental $U(1)^{N_f}$ flavor symmetry, where each $U(1)$ 
factor corresponds to a vector-like rotation of the quark fields of 
a given flavor.  The theory also possesses a non-anomalous $U(1)_R$ 
symmetry with
\begin{equation}
  R(Q) = R(\bar{Q}) = 1 - \frac{N_c}{N_f},
\qquad
  R(S) = \frac{2 N_c}{N_f}.
\label{eq:R-charges}
\end{equation}
To realize our setup, the $R$ symmetry must be spontaneously broken. 
Here we treat its effect through the background expectation value of 
the $S$ field, which is sufficient to understand the properties of 
dark matter.  We will eventually allow $S$ to be a dynamical field 
in Section~\ref{subsec:more_comp}.

Below the dynamical scale of QCD, which we take to be close to the 
scale charactering the entire supersymmetry breaking sector,%
\footnote{We discuss a potential difference between the dynamical 
 scale $\Lambda$ appearing here and the scale that breaks supersymmetry 
 in Section~\ref{subsec:more_comp}.}
the appropriate degrees of freedom to describe the dynamics are composite 
meson fields $M^i_{\bar{\jmath}} \sim Q^i \bar{Q}_{\bar{\jmath}}$ 
interacting via a non-perturbative superpotential~\cite{Intriligator:1995au}. 
Together with the couplings of Eq.~(\ref{eq:SQQ}), this leads to the 
effective superpotential
\begin{equation}
  W = \tilde{\lambda}_i \Lambda S M_i + \lambda_M (N_c - N_f) \Lambda^3 
      \left(\frac{\Lambda^{N_f}}{\det M}\right)^{\frac{1}{N_c - N_f}},
\label{eq:W_eff}
\end{equation}
where $\Lambda$ is the dynamical scale, and $M_i \equiv M^i_{\bar{\imath}}$ 
($i = \bar{\imath}$) are the diagonal mesons.  The coefficient $\lambda_M$ 
is an unknown factor coming from canonically normalizing the meson fields, 
and a similar factor for the first term is absorbed into the definition 
of $\tilde{\lambda}_i$.  Using naive dimensional analysis, the size of 
these coefficients are $\lambda_M \approx O((4\pi)^{-(2N_c-N_f)/(N_c-N_f)})$ 
and $\tilde{\lambda}_i/\lambda_i \approx O(1/4\pi)$.  Note that there 
are no low energy baryon fields when $N_f < N_c$.

Setting $\langle S \rangle \neq 0$ but $\langle F_S \rangle = 0$, 
the superpotential of Eq.~(\ref{eq:W_eff}) has a stable 
supersymmetry-preserving, $R$-violating minimum.  Defining 
$m_i = \tilde{\lambda}_i \langle S \rangle$ and $\det m = \prod_i m_i$, 
the minimum of the potential is at
\begin{equation}
  \langle M_i \rangle = \frac{\alpha_M \Lambda^2}{m_i},
\qquad
  \langle M^i_{\bar{\jmath}} \rangle = 0 
  \quad (i \neq \bar{\jmath}),
\label{eq:M_minimum}
\end{equation}
where for convenience we have defined $\alpha_M = \lambda_M 
(\det m/\lambda_M^{N_f} \Lambda^{N_f})^{1/N_c}$.   According to naive 
dimensional analysis, $m_i \approx O(\lambda \langle S \rangle/4\pi)$ and 
$\alpha_M \approx O((\lambda \langle S \rangle/\Lambda)^{N_f/N_c}/16\pi^2)$, 
where $\lambda$ represents the size of the original couplings in 
Eq.~(\ref{eq:SQQ}).  Since $\langle M^i_{\bar{\jmath}} \rangle = 0$ 
for $i \neq \bar{\jmath}$, the $U(1)^{N_f}$ flavor symmetry is unbroken, 
making the $M^i_{\bar{\jmath}}$ mesons with $i \neq \bar{\jmath}$ (for 
now, absolutely) stable.  Therefore, the lightest components of these 
fields, either scalars or fermions, are dark matter candidates.

The diagonal mesons $M_i$ are unstable.  Since the $R$ symmetry is 
assumed to be spontaneously broken, the theory has a light $R$~axion 
with the decay constant $f_a$ of order $\Lambda$.  The couplings 
of the mesons to the $R$~axion are determined by symmetries, and by 
doing appropriate field redefinitions, these couplings can be read 
off from the meson kinetic terms.%
\footnote{Once the $S$ field becomes dynamical, $S$ will mix with the 
 meson fields.  This can introduce extra contributions to the couplings, 
 especially if there is no hierarchy between the scales involved. 
 It, however, does not affect the basic conclusion here.}
For the scalar components, the kinetic term is ${\cal L} = |\partial_\mu 
M_{ii} + i c_M (\langle M_i \rangle + M_{ii}) (\partial_\mu a)/f_a|^2$, 
where $M_{ii} \equiv M_i - \langle M_i \rangle$ and $c_M$ is an $O(1)$ 
coefficient that depends on $R$ charges of the fields.  Assuming real 
$\langle M_i \rangle$, this gives a coupling of ${\rm Re}(M_{ii})$ to 
two $R$~axions, and coupling between ${\rm Re}(M_{ii})$, ${\rm Im}(M_{ii})$, 
and an $R$~axion.  Therefore, ${\rm Re}(M_{ii})$ decays promptly into two 
$R$~axions, while ${\rm Im}(M_{ii})$ decays (somewhat less) promptly into 
three $R$~axions via an off-shell ${\rm Re}(M_{ii})$.  After supersymmetry 
breaking, the fermionic component of $M_{ii}$ has a coupling to the 
corresponding scalar component and a gravitino, suppressed by powers 
of $\Lambda$.  This allows the $M_{ii}$ fermion to decay promptly 
into a gravitino and an $R$~axion.

\subsection{Thermal relic abundance}
\label{subsec:thermal}

Expanding around the minimum of the potential from Eq.~(\ref{eq:M_minimum}), 
the mass terms for the mesons are
\begin{equation}
  W =  \frac{\Lambda}{2 \alpha_M} \sum_{ij} 
    \left( M_{ij} M_{ji} \frac{m_i m_j}{\Lambda^2} 
      + \frac{1}{N_c-N_f} M_{ii} M_{jj} \frac{m_i m_j}{\Lambda^2} \right),
\label{eq:mass-terms}
\end{equation}
where we have used the notation $M_{ij} \equiv (M^i_{\bar{\jmath}} 
- \langle M^i_{\bar{\jmath}} \rangle)|_{\bar{\jmath}=j}$.  The stable 
mesons ($M_{ij}$ with $i \neq j$) do not have mixing terms because 
of the unbroken $U(1)^{N_f}$ symmetry, while the unstable mesons 
($M_{ii}$) have a mixing term that is suppressed by $1/(N_c - N_f)$. 
For simplicity of discussion, we will focus on the large $N_c$ limit 
where there is a simple relationship between the stable and unstable 
meson masses:
\begin{equation}
  \mbox{mass}(M_{ij}) \simeq 
    \sqrt{ \mbox{mass}(M_{ii})\, \mbox{mass}(M_{jj}) }.
\label{eq:approx-mass}
\end{equation}
Up to $1/(N_c - N_f)$ corrections, the leading interaction term for 
the mesons is
\begin{equation}
  W = -\frac{1}{3 \alpha_M^2} \sum_{ijk} 
      M_{ij} M_{jk} M_{ki} \frac{m_i m_j m_k}{\Lambda^3} 
    + \frac{1}{4 \alpha_M^3} \sum_{ijkl} 
      M_{ij} M_{jk} M_{kl} M_{li} \frac{m_i m_j m_k m_l}{\Lambda^5}.
\label{eq:triple-coupling}
\end{equation}

The superpotential coupling in Eq.~(\ref{eq:triple-coupling}) allows 
for various annihilation diagrams for stable $M_{ij}$ mesons.  By the 
mass relation in Eq.~(\ref{eq:approx-mass}), either the $ii$-type or 
$jj$-type mesons will be lighter than the $ij$-type mesons.  Assuming 
$m_i < m_j$, the annihilation cross section for $M_{ij} M_{ji} 
\rightarrow M_{ii} M_{ii}$ scales like
\begin{equation}
  \langle \sigma v \rangle \sim 
    \frac{1}{8 \pi} \frac{\kappa^4}{m_{\rm DM}^2},
\qquad
  m_{\rm DM} \sim \frac{1}{\alpha_M} \frac{m_i m_j}{\Lambda},
\qquad
  \kappa \sim \frac{1}{\alpha_M^2} \frac{m_i^2 m_j}{\Lambda^3}.
\label{eq:sigma-v-model}
\end{equation}
Using naive dimensional analysis, these quantities can be estimated as
\begin{equation}
  m_{\rm DM} \sim \Lambda\, \biggl( \frac{\lambda \langle S \rangle}{\Lambda} 
    \biggr)^{2-\frac{N_f}{N_c}},
\qquad
  \kappa \sim 4\pi\, \biggl( \frac{\lambda \langle S \rangle}{\Lambda} 
    \biggr)^{3-\frac{2 N_f}{N_c}},
\label{eq:NDA}
\end{equation}
so that the required annihilation rate can naturally be obtained.  For 
example, it is easy to obtain $m_{\rm DM} \approx O(10~{\rm TeV})$ and 
$\kappa$ a factor of a few, which leads to $\langle \sigma v \rangle 
\sim (1/8\pi)(1/{\rm TeV}^2)$, and thus $\Omega_{\rm DM} \simeq 0.2$. 
Note that here we have not included possible multiplicity factors in 
naive dimensional analysis, which will somewhat decrease the value 
of $\kappa$.  We therefore do not prefer very large $N_c$ in practice. 
Note also that, strictly speaking, our analysis is not theoretically 
very well under control for $\lambda \langle S \rangle \simgt \Lambda$, 
although we expect that the basic dynamics is still as presented for 
$\lambda \langle S \rangle \sim \Lambda$.

There is subtlety when annihilation occurs into states of comparable 
mass.  If $m_i \simeq m_j$, then the $ij$-, $ii$-, $jj$-type are nearly 
degenerate for relatively large $N_c$.  In this case, one must use the 
methods of Ref.~\cite{Griest:1990kh} to properly calculate the thermally 
averaged annihilation rate.  One can, however, still obtain the right 
thermal relic abundance with somewhat stronger couplings.  If $N_f$ 
is large, then there is a potential for multiple dark matter components 
having comparable abundances.  The freezeout calculation can then be 
affected by various co-annihilation channels.

\subsection{Dark matter decays}
\label{subsec:dm-decay}

The $U(1)^{N_f}$ flavor symmetry which ensures the stability of dark 
matter arises as an accidental symmetry of the renormalizable interactions 
of Eq.~(\ref{eq:SQQ}).  As such, it is plausible that this symmetry 
is not respected by physics at the unification or gravitational scale. 
Suppose that the leading operators encoding the high energy physics are 
the K\"{a}hler potential terms
\begin{equation}
  K = \frac{1}{M_*^2} \eta^{ik}_{jl} Q^\dagger_i Q^j Q^\dagger_k Q^l,
\label{eq:U1Nf-vio}
\end{equation}
with arbitrary flavor structures for $\eta^{ik}_{jl}$ in the 
basis where the interactions of Eq.~(\ref{eq:SQQ}) are diagonal: 
$\lambda_i^{\bar{\jmath}} = \lambda_i \delta_i^{\bar{\jmath}}$. 
Below the scale $\Lambda$, these operators can then be matched into%
\footnote{The same argument as below applies to the operator 
 $K = (1/M_*^2) \bar{Q}^{\dagger\bar{\imath}} \bar{Q}_{\bar{\jmath}} 
 \bar{Q}^{\dagger\bar{k}} \bar{Q}_{\bar{l}}$ or $(1/M_*^2) Q^\dagger_i 
 Q^j \bar{Q}^{\dagger\bar{k}} \bar{Q}_{\bar{l}}$.}
\begin{equation}
  K = \frac{\Lambda^2}{M_*^2} c^{i\bar{l}}_{\bar{\jmath}k} 
    M^{\dagger \bar{\jmath}}_i M^k_{\bar{l}},
\label{eq:U1Nf-vio_M}
\end{equation}
where $c^{i\bar{l}}_{\bar{\jmath}k}$ are coefficients.  Using 
naive dimensional analysis, we find $c^{i\bar{l}}_{\bar{\jmath}k} 
\approx O(1/16\pi^2)$ for $\eta^{ik}_{jl} \approx O(1)$ and 
$\lambda_i \langle S \rangle \approx O(\Lambda)$.  In general, 
one expects $c^{i\bar{l}}_{\bar{\jmath}k}$ to have $O(1)$ 
$CP$-violating phases.

The existence of the terms in Eq.~(\ref{eq:U1Nf-vio_M}) induces mixings 
between the diagonal and off-diagonal meson states, leading to the 
decay of dark matter.  The decay width has a large suppression from 
the $\Lambda^2/M_*^2$ factor in Eq.~(\ref{eq:U1Nf-vio_M}), so that the 
lifetime of dark matter is very long.  One possible decay chain arises 
through $M_{ii} \rightarrow aa$, where $a$ is the $R$~axion.  Through 
the mixing, this leads to
\begin{equation}
  M_{ij} \rightarrow aa,
\label{eq:DM-decay-aa}
\end{equation}
where we have assumed that dark matter $M_{ij}$ is a scalar.%
\footnote{Because of the $CP$-violating phases in 
 $c^{i\bar{l}}_{\bar{\jmath}k}$, even ${\rm Im}(M_{ij})$ 
 dominantly decays into two $R$~axions, not three.}
For fermionic dark matter, one of the $a$'s must be replaced by a gravitino 
$\tilde{G}$.  Another possibility is that dark matter decays through 
the meson self interactions of Eq.~(\ref{eq:triple-coupling}).  Again, 
through the mixing, this can lead to
\begin{equation}
  M_{ij} \rightarrow M_{ii} M_{ii},
\label{eq:DM-decay-MM}
\end{equation}
if it is kinematically allowed, ${\rm mass}(M_{ij}) > 2\, 
{\rm mass}(M_{ii})$.  Here, $M_{ii}$ in the final state can 
be either a scalar or fermion, which subsequently decays into 
$aa$ or $a\tilde{G}$, respectively.

For the scalar dark matter decay in Eq.~(\ref{eq:DM-decay-aa}), the 
lifetime of dark matter can be estimated as
\begin{equation}
  \tau_{\rm DM} \approx 8\pi \frac{f_a^4}{m_{\rm DM}^3 \langle M \rangle^2} 
    \biggl( \frac{M_*^2}{c \Lambda^2} \biggr)^2 
  \simeq 3 \times 10^{27}~{\rm sec}\, 
    \biggl( \frac{1/16\pi^2}{c} \biggr)^2 
    \left( \frac{M_*}{10^{17}~{\rm GeV}} \right)^4 
    \left( \frac{10~{\rm TeV}}{\Lambda} \right)^5,
\label{eq:tau-model}
\end{equation}
where $\langle M \rangle$ is a typical meson expectation value, $c$ 
represents a generic size for the coefficients $c^{i\bar{l}}_{\bar{\jmath}k}$ 
in Eq.~(\ref{eq:U1Nf-vio_M}), and we have used $m_{\rm DM} \approx 4\pi f_a 
\approx \Lambda$ in the last equation.  For $M_*$ of order the unification 
scale, the lifetime is in the range required to produce observable cosmic 
ray signatures.  While there are different parametric dependences for 
$\tau_{\rm DM}$ if one uses the decay mode in Eq.~(\ref{eq:DM-decay-MM}) 
or if one considers fermionic dark matter decay, they all give the same 
estimate as Eq.~(\ref{eq:tau-model}) after using naive dimensional analysis.

\subsection{Towards a more complete theory}
\label{subsec:more_comp}

In the above discussion, we assumed that $\langle S \rangle$ could be 
treated as a spurion for the spontaneous breaking of the $R$ symmetry. 
In a complete theory, $S$ would be a propagating degree of freedom 
that is part of a larger supersymmetry breaking sector.  Naively 
coupling $S$ to the mesons as in Eq.~(\ref{eq:W_eff}), the $F_S$-term 
potential would change the meson vacuum structure.  In particular, even 
if $S$ could be stabilized through terms in the K\"{a}hler potential, 
there is generically a mesonic runaway direction~\cite{Izawa:2009mj}.

Therefore, it is important to have a proof-of-concept that a propagating 
$S$ field can not only obtain an $R$-violating vacuum expectation value 
but also couple to the mesons without introducing runaway behavior. 
In addition, one would like to see that appropriate messenger fields can 
be added to the theory to communicate supersymmetry breaking to the SSM. 
While a complete theory would incorporate these two effects in a completely 
dynamical setting with all scales set by dimensional transmutation, 
here we consider O'Raifeartaigh-type models in order to treat these 
effects modularly (with a dynamical model mentioned only briefly towards 
the end).  We leave further model building to future work, although we 
emphasize that the precise details of supersymmetry breaking are largely 
irrelevant for the dark matter discussion.

We will also not be concerned by the specific mass hierarchies needed 
to realize a realistic theory.  Typically, the scale of supersymmetry 
breaking should be closer to $O(100~{\rm TeV})$ to obtain a realistic 
superparticle spectrum, while the mass scale for dark matter suggested 
by the cosmic ray data is closer to $O(10~{\rm TeV})$.  In a single 
scale theory with naive dimensional analysis, both mass scales are 
expected to coincide.  However, in a complete dynamical theory of 
supersymmetry breaking, there may be additional structures that 
generate such a ``little'' hierarchy.%
\footnote{This little hierarchy may not be necessary in general if 
 the axion-like state decays into $\pi^+\pi^-\pi^0$ or $\tau^+\tau^-$ 
 (see Figure~\ref{fig:best-fit}) and the number of messenger fields 
 is relatively large.}
For example, this can be realized if the $SU(N_c)$ theory in 
Section~\ref{subsec:setup} is in a strongly interacting conformal 
window at high energies.  If nontrivial dynamics kicks in at the 
scale $O(100~{\rm TeV})$ to break supersymmetry, the same dynamics 
can make the $SU(N_c)$ theory deviate from the fixed point, presumably 
due to decoupling of some degrees of freedom, leading to the dark 
matter setup considered here.  The physics of $R$ breaking will 
be associated with the higher scale dynamics.

To see how $S$ can be made dynamical, consider adding the following 
interactions to Eq.~(\ref{eq:W_eff}):
\begin{equation}
  W = \lambda_X X \left(S \bar{S} - \mu_S^2 \right) 
    + m_S S Y + \bar{m}_S \bar{S} \bar{Y}.
\label{eq:OR}
\end{equation}
These interactions could be the low energy description of a more complete 
supersymmetry breaking sector, and so we consider that $\mu_S$, $m_S$, 
and $\bar{m}_S$ are roughly of order $\Lambda$.  Regardless of the $R$ 
charge of the $S$ field (i.e.\ regardless of $N_f$ and $N_c$), there 
is a unique consistent $R$ charge assignment for the new fields, with
\begin{equation}
  R(X) = 2,
\qquad
  R(\bar{S}) = -\frac{2 N_c}{N_f},
\qquad
  R(Y) = 2 - \frac{2 N_c}{N_f},
\qquad
  R(\bar{Y}) = 2 + \frac{2 N_c}{N_f}.
\label{eq:R-OR}
\end{equation}
With the inclusion of the $m_S$ term, the $F_S$ equation of motion 
simply sets the value of $Y$ and does not change the potential for 
the mesons, so that the global minimum of the meson potential is 
still given by Eq.~(\ref{eq:M_minimum}).  The $\lambda_X$ coupling 
forces $S$ and $\bar{S}$ to obtain vacuum expectation values, and 
the $m_S$ and $\bar{m}_S$ terms constrain $S$ and $\bar{S}$ from 
running away to infinity.  As long as $\lambda_X^2 \mu_S^2 > m_S 
\bar{m}_S$, this theory has a supersymmetry- and $R$-breaking 
global minimum away from the origin, with $S$ vacuum expectation 
value given by
\begin{equation}
  \langle S \rangle = \sqrt{\frac{\bar{m}_S}{m_S} 
    \left( \mu_S^2 - \frac{m_S \bar{m}_S}{\lambda_X^2} \right)}.
\label{eq:S_vev}
\end{equation}

At tree level, the only fields that obtain $F$-term vacuum expectation 
values are $X$, $Y$, and $\bar{Y}$, so the off-diagonal mesons only 
feel supersymmetry breaking at loop level.  The vacuum expectation 
value of $X$ is unspecified at tree level, but one expects loop effects 
to stabilize $X$, though not necessarily at $\langle X \rangle = 0$. 
Since supersymmetry and $R$ are both broken regardless of the value of 
$\langle X \rangle$, this is an example of a model where supersymmetry 
and $R$ are spontaneously broken at tree level.%
\footnote{In the language of Ref.~\cite{Komargodski:2009jf}, 
 Eq.~(\ref{eq:OR}) is the ``$g = 0$'' superpotential which preserves 
 a $U(1)_R$ symmetry and spontaneously breaks a global $U(1)_A$ symmetry. 
 The inclusion of Eq.~(\ref{eq:SQQ}) identifies a linear combination 
 of $U(1)_R$ and $U(1)_A$ as the true $U(1)'_R$ symmetry, such that 
 the $R$ symmetry is broken.}

In principle, one could calculate the full spectrum of the theory at 
one loop, allowing one to test whether the stable dark matter is the 
fermionic or bosonic component of the off-diagonal mesons.  In practice, 
this is a nontrivial exercise given the large amount of mixing between 
the various new degrees of freedom.  If the dynamical sector is truly 
strongly coupled, then loop counting would not apply, and the mesons 
could feel $O(1)$ supersymmetry breaking effects depending on the 
details of the K\"{a}hler potential.  Therefore, one would not expect 
a one-loop calculation to get the correct sign for the fermion/boson 
splitting.

We have seen that Eq.~(\ref{eq:SQQ}) can be consistently coupled to 
the supersymmetry- and $R$-breaking sector in Eq.~(\ref{eq:OR}).  It 
is then straightforward to couple messenger fields to the supersymmetry 
breaking, using an analogous structure to Eq.~(\ref{eq:OR}).  Regardless 
of how $X$ is stabilized, the $Y$ field obtains vacuum expectation 
values in both the lowest and highest components, so $Y$ can communicate 
supersymmetry breaking to the SSM sector through complete $SU(5)_{\rm SM}$ 
messenger multiplets.  Consider adding the superpotential
\begin{equation}
  W = \lambda_Y Y F \bar{F} + m_F F \bar{F}' + \bar{m}_F F' \bar{F},
\label{eq:W_mess}
\end{equation}
where $F$ and $F'$ are ${\bf 5}$s of $SU(5)_{\rm SM}$, and $\bar{F}$ 
and $\bar{F}'$ are ${\bf \bar{5}}$s, and there is a consistent $R$ 
charge assignment for the messengers.  Taking $m_F \bar{m}_F > \lambda_Y 
m_S \langle S \rangle$, we can ensure that the messengers do not develop 
vacuum expectation values.  With this kind of messenger sector, the gaugino 
masses are proportional to $F_Y^3$ instead of $F_Y$~\cite{Izawa:1997gs}. 
They are, therefore, suppressed compared with the scalar masses for 
$\lambda_Y F_Y \ll m_F \bar{m}_F$, which may not be fully desired. 
Moreover, one also needs to take $m_F \simeq \bar{m}_F$ to avoid 
an unwanted large Fayet-Iliopoulos term for $U(1)$ hypercharge.  As 
a proof-of-concept, however, we see that it is possible to couple 
the dark matter, supersymmetry breaking, and messenger sectors 
together in a consistent way.

If one desires, the above model can be extended in such a way that 
all the dimensionful scales $\mu_S$, $m_S$, $\bar{m}_S$, $m_F$, and 
$\bar{m}_F$ are generated dynamically.  These scales can be replaced 
by a single chiral superfield $T$, which obtains a vacuum expectation 
value through $W = Z (T^2 - \mu_T^2)$.  The new scale $\mu_T$ can then 
be generated from dimensional transmutation, e.g., by replacing $\mu_T^2$ 
with a quark condensate as in Ref.~\cite{Izawa:1997gs}.  For appropriate 
parameter choices, this extension leaves the O'Raifeartaigh dynamics 
largely intact.  To connect the $\mu_T$ scale to the $\Lambda$ scale 
of supersymmetric QCD, we can introduce extra quarks $Q'$ and $\bar{Q}'$ 
of $SU(N_c)$ that make the theory conformal at high energies.  The 
quarks obtain masses from $W = T Q' \bar{Q}'$, triggering the exit 
from a strongly-coupled fixed point.  This provides an existence-proof 
model where all the scales are generated dynamically associated with 
single dimensional transmutation.  The $\mu$ term can also be generated 
by introducing singlet ``messengers'' and coupling them, together 
with the doublet messengers, to the SSM Higgs fields as in 
Ref.~\cite{Csaki:2008sr}.

Note that there is a $Z_2$ parity in Eq.~(\ref{eq:W_mess}) under which 
all the messenger fields are odd.  Such a parity is often present 
in realistic constructions for the supersymmetry breaking sector. 
In order to avoid the problem of unwanted colored/charged relics, 
however, the messenger fields (either elementary or composite) must 
decay.  If the decay occurs through dimension five operators suppressed 
by $M_*$ and if the decay products contain an SSM state, then this may 
explain the discrepancy of the measured ${}^7{\rm Li}$ abundance from 
the prediction of standard big-bang nucleosynthesis, along the lines 
of Ref.~\cite{Arvanitaki:2008hq}.  In fact, the lowest dimension for 
the operators causing messenger decay can easily be five if, for example, 
the messenger fields are two-body bound states, such as meson states, 
of some strong dynamics.

\section{Astrophysical Signatures}
\label{sec:astro}

We have seen that dark matter in the present scenario naturally has 
the following features, which are not shared by the standard weakly 
interacting massive particle (WIMP) scenario:
\begin{itemize}
\item
The mass of dark matter is of $O(10~{\rm TeV})$, which is significantly 
larger than the weak scale.  The correct thermal abundance, however, is 
still obtained because of the relatively large annihilation cross section.
\item
Dark matter can decay through dimension six operators, and thus with 
lifetime of $O(10^{26}~{\rm sec})$.  This can lead to observable 
cosmic ray signatures.
\item
Dark matter can decay into light axion-like states (and to the gravitino 
if dark matter is a fermion), which in turn decays into standard model 
particles.  The final states can naturally be only $e^+e^-$, $\mu^+\mu^-$, 
$\pi^+\pi^-\pi^0$, or $\tau^+\tau^-$.
\end{itemize}
In fact, these features are precisely what are needed to explain recent 
electron/positron cosmic ray data from PAMELA, FERMI, and H.E.S.S. 
In this section, we demonstrate that recent astrophysical data are 
indeed beautifully explained in the present setup, and provide a fit 
to the parameters $m_{\rm DM}$ and $\tau_{\rm DM}$ using the observed 
electron/positron fluxes.  We also discuss the diffuse $\gamma$-ray 
flux that could be seen in the near future by experiments such as FERMI.

Let us begin with a summary of the observational situation.  The PAMELA 
experiment has recently reported an unexpected rise in the positron 
fraction $\Phi_{e^+}/(\Phi_{e^+}+\Phi_{e^-})$ in the energy range between 
about $10$ and $100~{\rm GeV}$~\cite{Adriani:2008zr}.  On the other 
hand, they did not see any deviation from the expected background in 
the antiproton data~\cite{Adriani:2008zq}.  The FERMI experiment is 
the first to be able to measure the combined electron and positron 
flux with good precision and control of uncertainties over the entire 
range from $20~{\rm GeV}$ to $1~{\rm TeV}$.  Their recent data also show 
an excess over standard diffuse cosmic ray backgrounds~\cite{Abdo:2009zk}, 
with a broad structure extending up to the highest energies.%
\footnote{The ATIC~\cite{Chang:2008zzr} and PPB-BETS~\cite{Torii:2008xu} 
 experiments have also reported an excess in the combined flux at 
 around $600~{\rm GeV}$, although with a peaked spectral shape that does 
 not seem to fully agree with the FERMI data.  With their significant 
 statistical and experimental uncertainties, we do not include these 
 results in our analysis.}
At higher energies, data from the H.E.S.S. experiment indicates 
a spectral break in the combined electron and positron flux at around 
$1~{\rm TeV}$, with the spectral index increasing from $\approx 3.0$ 
to $\approx 4.1$~\cite{Aharonian:2009ah}.

While there remain experimental uncertainties as well as difficulty 
in calculating astrophysical background fluxes, taken together these 
results suggest a new source of primary electrons and positrons with 
a broad spectrum extending up to a few TeV.  An exciting possibility 
is that these are signals of annihilation or decay of galactic dark 
matter~\cite{Cirelli:2008pk,Nardi:2008ix}, although astrophysical 
interpretations, e.g.\ in terms of nearby pulsars~\cite{Profumo:2008ms}, 
are also possible.  When interpreted in terms of dark matter, the 
data has certain implications:
\begin{itemize}
\item
The structure around TeV in the combined $e^+$ and $e^-$ flux is very 
broad, implying that electrons/positrons do not arise directly from dark 
matter annihilation or decay; rather, they arise through some cascading 
processes.
\item
The spectral cutoff around TeV, together with the broad structure, 
implies that the mass of dark matter is larger than TeV.  In particular, 
if cosmic rays arise from decay of dark matter, then the mass scale 
is more like $O(10~{\rm TeV})$, especially if the cascade is sufficiently 
long.
\item
The absence of signal in the antiproton data implies that dark matter 
annihilates or decays mainly into leptons, although precisely how much 
nucleonic final states should be suppressed is not completely clear 
because of uncertainties in proton/antiproton propagation models.
\end{itemize}
These interpretations are consistent with detailed analyses performed 
after the recent FERMI data release~\cite{Bergstrom:2009fa,Meade:2009iu}.

The fact that a naive WIMP dark matter candidate, such as the neutralino 
lightest supersymmetric particle (LSP), does not satisfy the above 
criteria is quite suggestive.  In the present scenario, dark matter 
mass is naturally of $O(10~{\rm TeV})$, which decays with lifetime 
of $O(10^{26}~{\rm sec})$ producing the observed $e^\pm$ signatures. 
The decay occurs through a light axion-like state (axion portal), 
so that the final states can selectively be leptons.%
\footnote{An $R$~axion as the portal was mentioned in 
 Ref.~\cite{Nomura:2008ru} and developed further in 
 Ref.~\cite{Ibe:2009dx} in the context of annihilating 
 dark matter.  For a related suggestion in dark matter annihilation, 
 see~\cite{Banks:2009rb}.}
This also ensures that the final state leptons arise through cascades, 
making the spectrum consistent with the latest FERMI and H.E.S.S. 
data.  It is also worth mentioning that decaying dark matter is 
much less constrained than annihilating dark matter~\cite{Meade:2009iu}, 
which has some tension with $\gamma$-ray and neutrino 
observations~\cite{Bertone:2008xr,Mardon:2009rc}.

\subsection{PAMELA, FERMI, and H.E.S.S. electron/positron data}
\label{subsec:e-signals}

There are various possible decay chains in the present scenario, 
depending on which state is dark matter and the decay properties of 
the axion-like state $a$.  Here we consider two representative classes: 
$1$-step cascade ($\phi \rightarrow aa$; $a \rightarrow \ell^+ \ell^-$) 
and $2$-step cascade ($\phi \rightarrow \phi'\phi'$; $\phi' \rightarrow 
aa$; $a \rightarrow \ell^+ \ell^-$), where $\phi$ and $\phi'$ represent 
dark matter and another state in the supersymmetry breaking sector, 
respectively, and $\ell = e, \mu, \tau$.  We also consider $a \rightarrow 
\pi^+\pi^-\pi^0$.  An example with a $1$-step cascade was already seen 
in Section~\ref{sec:illust}: dark matter is (the scalar component of) 
the lightest ``meson,'' which decays into two axion-like states.  The 
axion-like state then decays into standard model fields, as seen in 
Section~\ref{subsec:axi-const}.  A $2$-step cascade can easily arise 
if (the scalar component of) the lightest ``baryon'' is dark matter. 
After including dimension six baryon number violating operators, this 
state can decay into two meson states, each of which then decays into 
two axion-like states.  Alternatively, dark matter may be a meson 
state which dominantly decays into other meson states, as seen in 
Section~\ref{subsec:dm-decay}.

\begin{figure}[t]
\begin{center}
\begin{picture}(280,120)(195,-10)
%
%
  \Text(220,-8)[t]{\large $1$ step}
  \Line(160,50.2)(200,50.2) \Line(160,49.8)(200,49.8)
  \Text(157,50)[r]{$\phi$}
  \Line(200,50)(240,70) \Text(225,66)[br]{$a$}
  \Line(200,50)(240,30) \Text(225,35)[tr]{$a$}
  \Line(240,70)(280,80) \Text(285,82)[l]{$\ell^+$}
  \Line(240,70)(280,60) \Text(285,60)[l]{$\ell^-$}
  \Line(240,30)(280,40) \Text(285,42)[l]{$\ell^+$}
  \Line(240,30)(280,20) \Text(285,20)[l]{$\ell^-$}
%
%
  \Text(410,-8)[t]{\large $2$ step}
  \Line(350,50.2)(380,50.2) \Line(350,49.8)(380,49.8)
  \Text(347,50)[r]{$\phi$}
  \Line(380,50)(410,70) \Text(401,64)[br]{$\phi'$}
  \Line(380,50)(410,30) \Text(401,34)[tr]{$\phi'$}
  \Line(410,70)(440,85) \Text(431,83)[br]{$a$}
  \Line(410,70)(440,60) \Text(431,61)[tr]{$a$}
  \Line(410,30)(440,40) \Text(431,39)[br]{$a$}
  \Line(410,30)(440,15) \Text(431,18)[tr]{$a$}
  \Line(440,85)(470,90) \Text(475,93)[l]{$\ell^+$}
  \Line(440,85)(470,80) \Text(475,81)[l]{$\ell^-$}
  \Line(440,60)(470,65) \Text(475,68)[l]{$\ell^+$}
  \Line(440,60)(470,55) \Text(475,56)[l]{$\ell^-$}
  \Line(440,40)(470,45) \Text(475,46)[l]{$\ell^+$}
  \Line(440,40)(470,35) \Text(475,34)[l]{$\ell^-$}
  \Line(440,15)(470,20) \Text(475,21)[l]{$\ell^+$}
  \Line(440,15)(470,10) \Text(475,9)[l]{$\ell^-$}
\end{picture}
\end{center}
\caption{Cascade decays of dark matter $\phi$ through an axion-like 
 state $a$.  Here, $\phi'$ is an unstable state in the supersymmetry 
 breaking sector, and $\ell^\pm$ ($\ell = e,\mu,\tau$) are standard 
 model leptons.}
\label{fig:cascade}
\end{figure}
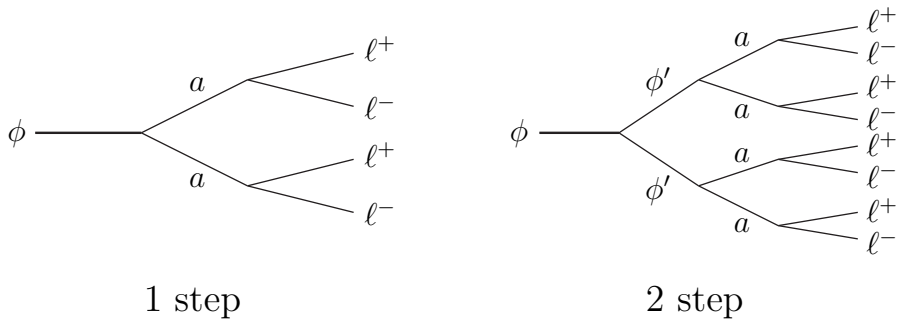
The processes just described are depicted in Figure~\ref{fig:cascade}. 
In the case where dark matter is a fermionic component, one side of 
some step of the cascade should be replaced by the gravitino, but this 
does not change the predicted signals, except that the dark matter 
lifetime must be rescaled by appropriate factors.  Below, we present 
the fit of the predicted $e^\pm$ fluxes to the PAMELA and FERMI data. 
We find that good agreements between the predictions and data are 
obtained for all the cases considered, except that the $1$-step 
$e^\pm$ case may have some tension with the H.E.S.S. data.  While 
we only present the results for $a \rightarrow e^+e^-$, $\mu^+\mu^-$ 
and $\tau^+\tau^-$ here, we also performed the same analyses for 
$a \rightarrow \pi^+\pi^-\pi^0$.  We find that the results in this 
case are very similar to the case of $a \rightarrow \tau^+\tau^-$, 
since the final state $e^\pm$ spectrum arising from charged pion 
decay is similar to that arising from $\tau$ decay.

Our analysis here follows that of Ref.~\cite{Mardon:2009rc}, where 
galactic propagation of $e^\pm$ is treated by the standard diffusion-loss 
equation.  The primary injection spectra are calculated as described 
there, assuming a large mass hierarchy in each cascade step.  In addition, 
tau decays are simulated using \textsc{Pythia}~8.1~\cite{Sjostrand:2000wi}. 
We assume an NFW halo profile~\cite{Navarro:1996gj} and use the MED 
propagation model parametrization given in~\cite{Delahaye:2007fr}. 
While there is significant uncertainty in the halo profile within 
a few kpc of the galactic center, its effect on the predicted 
$e^\pm$ fluxes is small.  We treat the astrophysical background as in 
Refs.~\cite{Mardon:2009rc,Cirelli:2008pk}: we take the parameterization 
of~\cite{Moskalenko:1997gh} and marginalize over the overall slope 
$P$ and normalization $A$ in the range $-0.05 < P_{e^+,e^-} < 0.05$ 
and $0 < A_{e^+,e^-} < \infty$.  There remain substantial uncertainties 
in the background at the high energies explored by H.E.S.S., where 
primary electron fluxes depend strongly on individual sources within 
$\approx 1~{\rm kpc}$ from the Earth.  We therefore do not include 
the H.E.S.S. data in our fit procedure.

\begin{figure}
  \center{\includegraphics[scale=0.9]{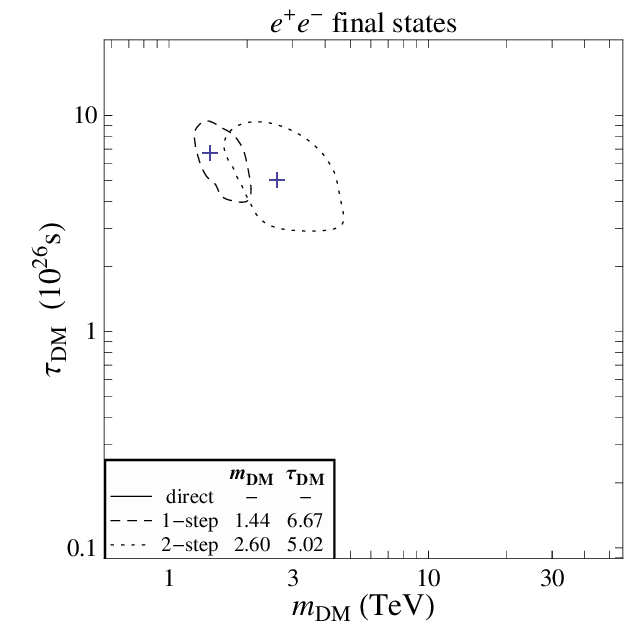}
  \includegraphics[scale=0.9]{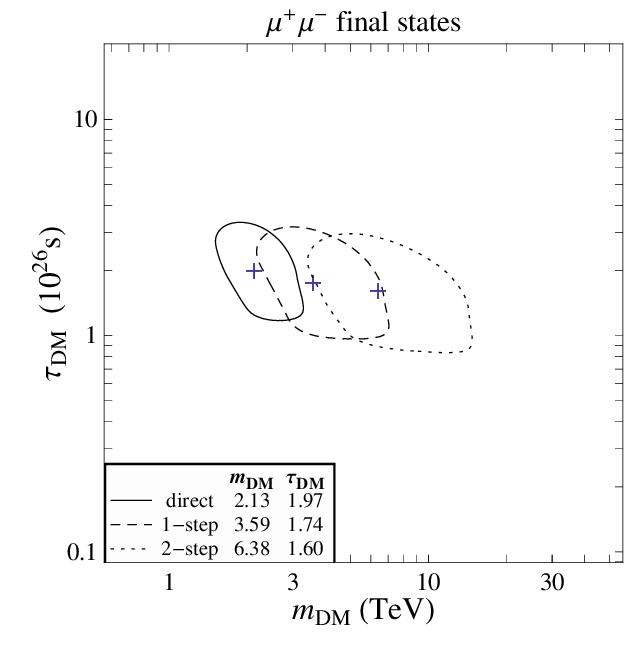}
  \includegraphics[scale=0.9]{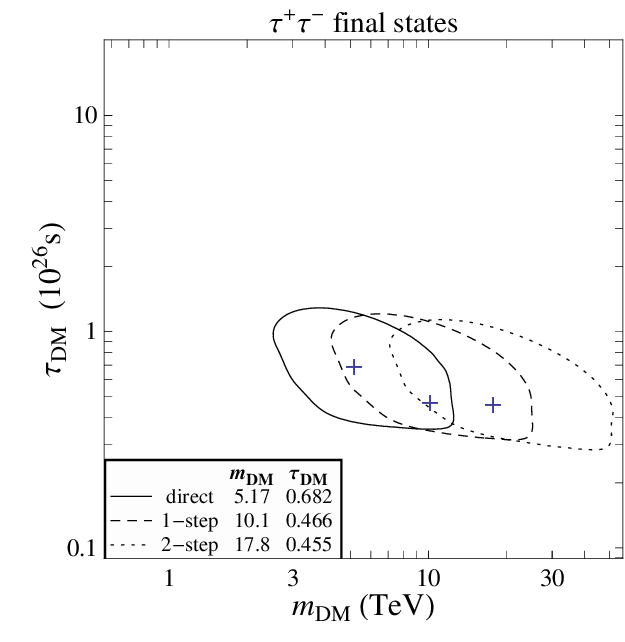}}
\caption{Regions of best fit (at $68\%$~C.L.) to the PAMELA and FERMI 
 data for dark matter mass $m_{\rm DM}$ and lifetime $\tau_{\rm DM}$, 
 in the case of direct (solid), $1$-step (dashed), and $2$-step (dotted) 
 decays into $e^+e^-$, $\mu^+\mu^-$, and $\tau^+\tau^-$.  The best fit 
 values of $m_{\rm DM}$ and $\tau_{\rm DM}$ are indicated by the crosses, 
 and are displayed inset in units of TeV and $10^{26}~{\rm sec}$, 
 respectively.  Direct decays into $e^+e^-$ does not give a good fit. 
 The case of $\pi^+\pi^-\pi^0$ is similar to that of $\tau^+\tau^-$.}
\label{fig:best-fit}
\end{figure}

We performed a $\chi^2$ analysis of signal plus background fluxes 
to the PAMELA and FERMI data.  The PAMELA data at energies less than 
$10~{\rm GeV}$ is strongly affected by solar modulation, and we exclude 
it from our analysis.  The FERMI experiment released both statistical 
and systematic errors with their data.  We conservatively combine 
these in quadrature, but note that this is likely an overestimation 
of the errors.  The FERMI data is also subject to an overall systematic 
uncertainty in energy of ${}^{+5\%}_{-10\%}$, under which all energies 
are rescaled as $E \to r E$; we therefore also marginalize over $r$ 
in the range $0.9 < r < 1.05$.  The result for best fit regions of 
dark matter mass, $m_{\rm DM}$, and lifetime, $\tau_{\rm DM}$ is 
shown in Figure~\ref{fig:best-fit}, for $1$-step and $2$-step cascades 
to electrons, muons and taus.
For comparison, we also show the fits for direct decays.  With $26$ 
degrees of freedom ($7~\mbox{PAMELA} + 26~\mbox{FERMI} - 7~\mbox{fitting 
parameters}$), we plot $68\%$~CL contours, corresponding to $\chi^2 
= 28.8$.  The best-fit values of $m_{\rm DM}$ and $\tau_{\rm DM}$ 
are indicated in each case.

\begin{figure}
  \center{\includegraphics[scale=0.98]{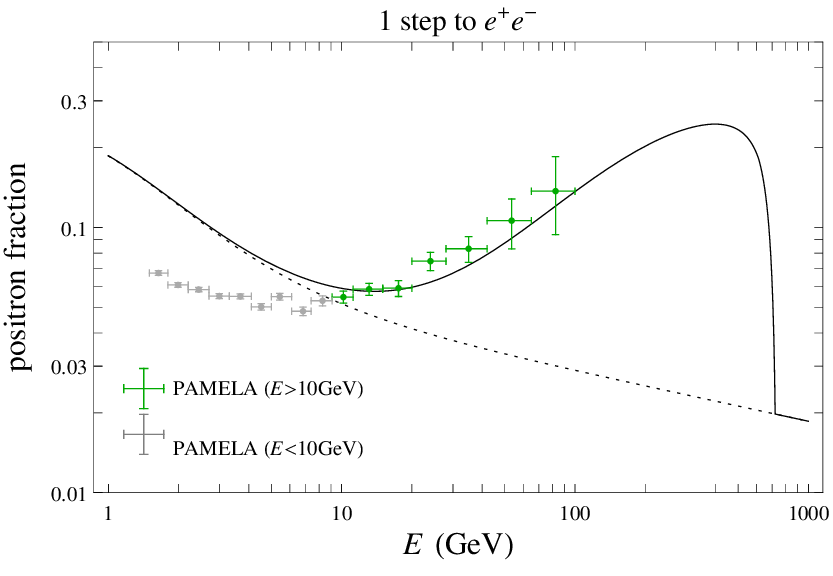}
  \hspace{0.5cm}
  \includegraphics[scale=0.98]{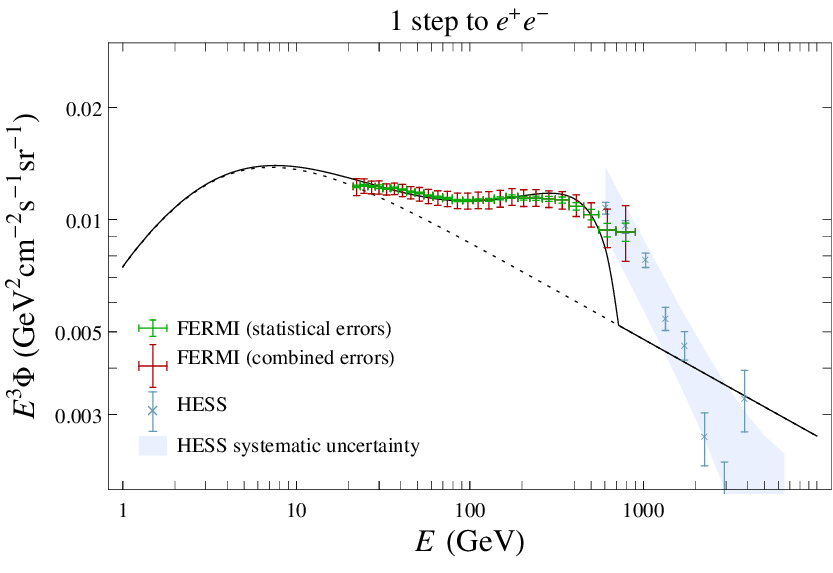}}
  \center{\includegraphics[scale=0.98]{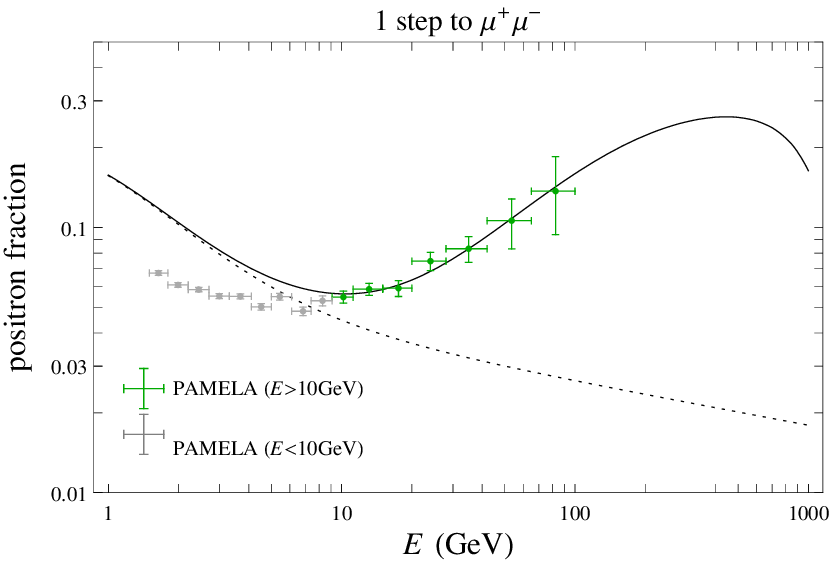}
  \hspace{0.5cm}
  \includegraphics[scale=0.98]{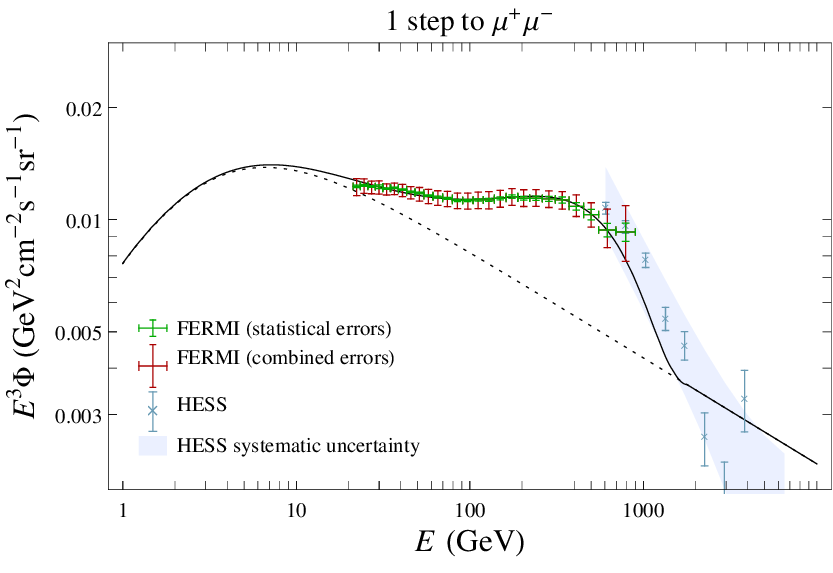}}
  \center{\includegraphics[scale=0.98]{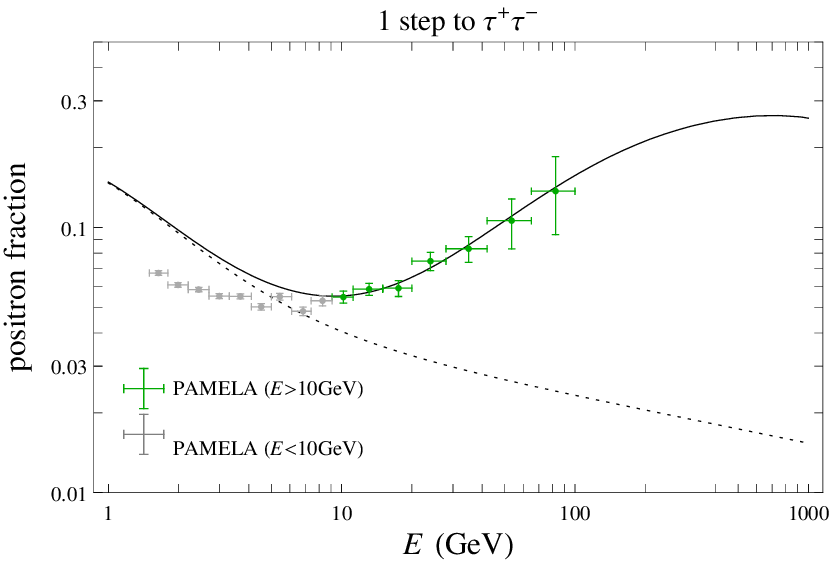}
  \hspace{0.5cm}
  \includegraphics[scale=0.98]{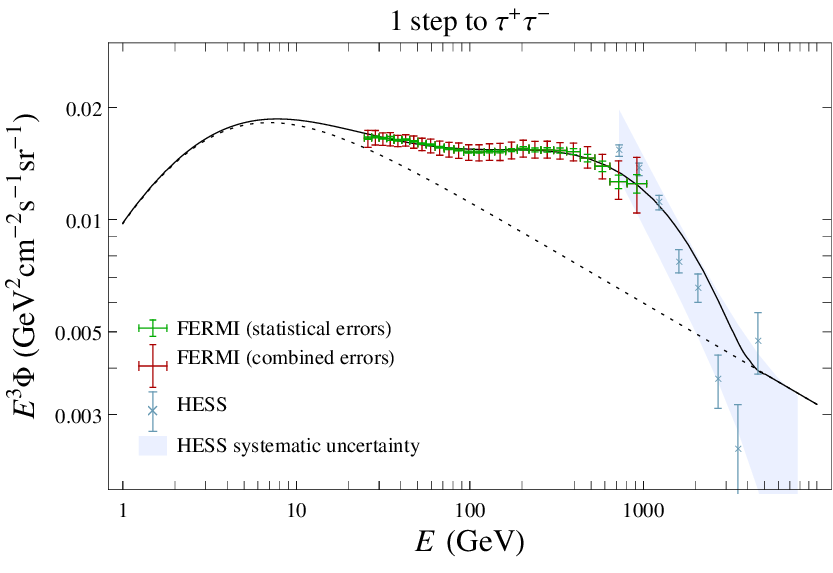}}
\caption{The predicted $e^\pm$ fluxes compared to the PAMELA and FERMI 
 data for $1$-step cascade decays into $e^+e^-$, $\mu^+\mu^-$, and 
 $\tau^+\tau^-$.  In each case, the mass and lifetime of dark matter 
 are chosen at the best fit point indicated in Figure~\ref{fig:best-fit}, 
 with the background (dotted) and FERMI energy-normalization marginalized 
 as described in the text.  We overlay the H.E.S.S. data with energy 
 rescaled in the range $\pm 15\%$ to best match the theory.  Note that 
 due to considerable uncertainty in the background fluxes at H.E.S.S. 
 energies, direct comparison of predicted fluxes with the H.E.S.S. 
 data may be misleading.}
\label{fig:1-step}
\end{figure}

\begin{figure}
  \center{\includegraphics[scale=0.98]{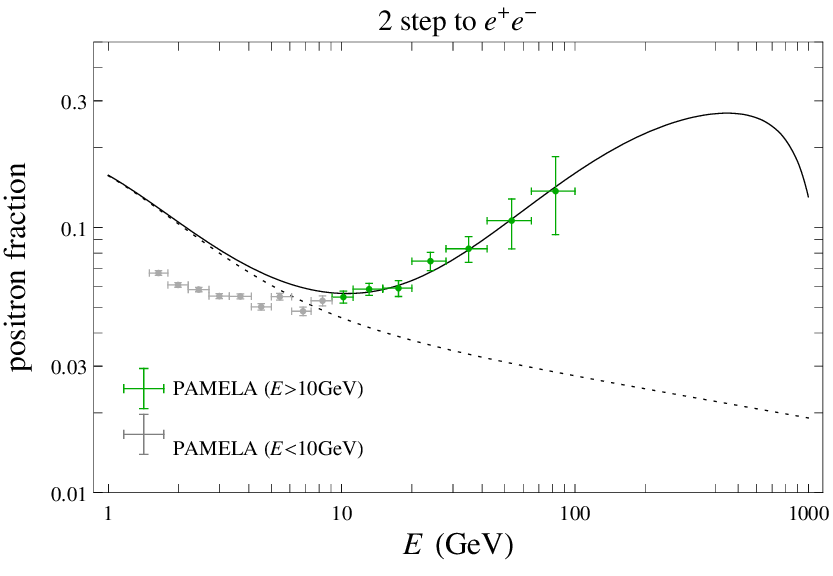}
  \hspace{0.5cm}
  \includegraphics[scale=0.98]{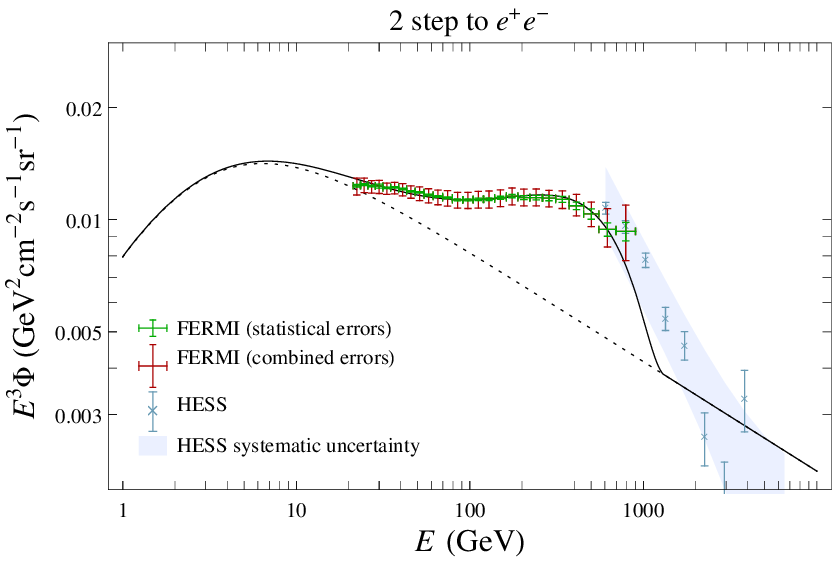}}
  \center{\includegraphics[scale=0.98]{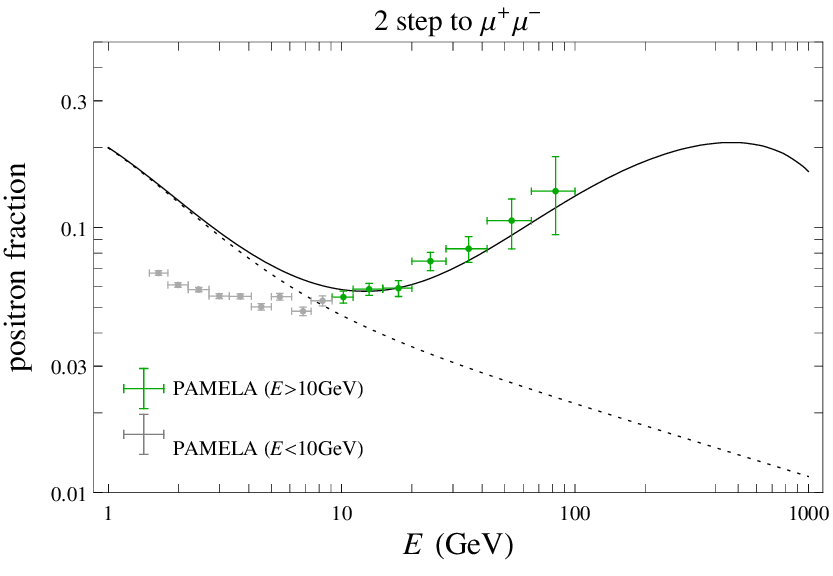}
  \hspace{0.5cm}
  \includegraphics[scale=0.98]{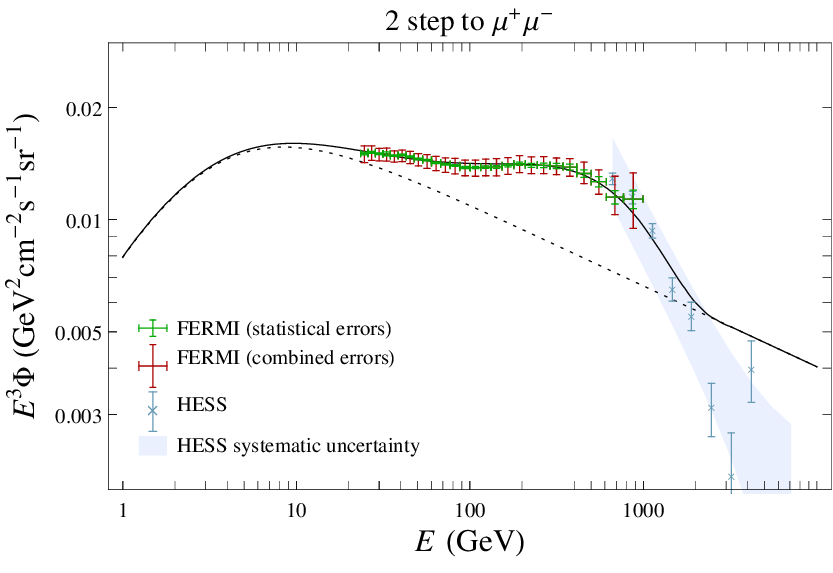}}
  \center{\includegraphics[scale=0.98]{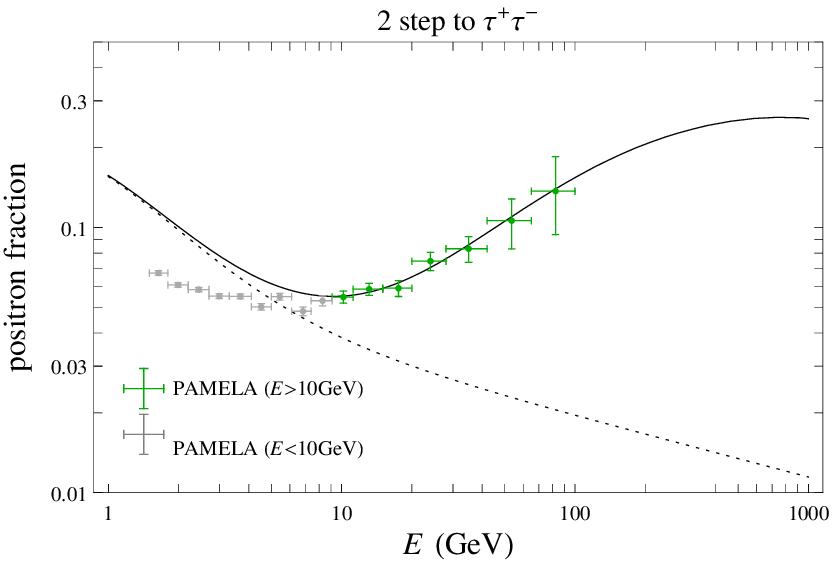}
  \hspace{0.5cm}
  \includegraphics[scale=0.98]{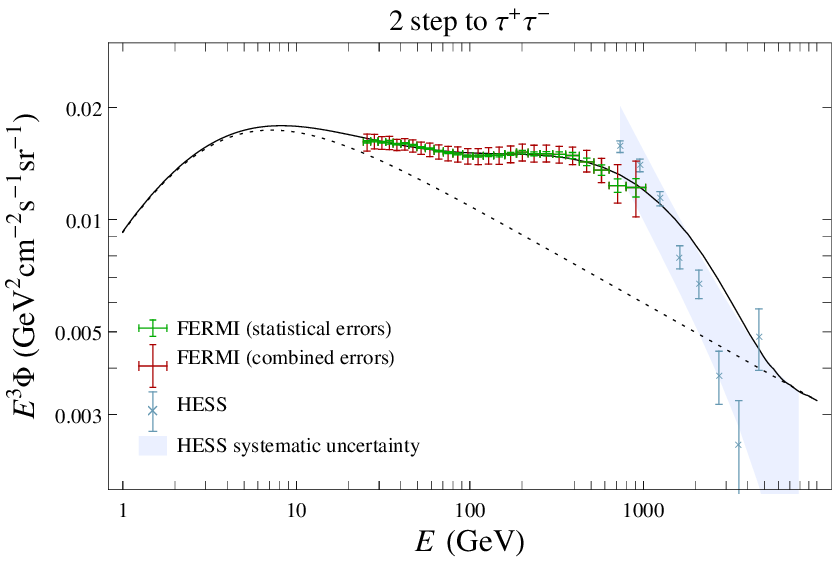}}
\caption{The same as Figure~\ref{fig:1-step} for $2$-step cascade decays 
 into $e^+e^-$, $\mu^+\mu^-$, and $\tau^+\tau^-$.}
\label{fig:2-step}
\end{figure}

We find that good fits are obtained in the region $m_{\rm DM} 
\approx O(1~\mbox{--}~100~{\rm TeV})$ and $\tau_{\rm DM} \approx 
O(10^{25}~\mbox{--}~10^{27}~{\rm sec})$, depending on the decay chain. 
In Figs.~\ref{fig:1-step} and \ref{fig:2-step} we show the predicted 
$e^\pm$ fluxes compared to the PAMELA and FERMI data, using the best 
fit parameters obtained from the $\chi^2$ analysis.
The H.E.S.S. high energy data~\cite{Aharonian:2008aaa} is also overlaid, 
with the energy rescaled in each case to best match the predicted 
flux, within the $\pm 15\%$ range of overall systematic uncertainty 
in H.E.S.S. energy.  The agreement between the predictions and data 
is remarkable.  We find some tension with the H.E.S.S. data in the case 
of $1$-step to $e^+e^-$, but in all other cases the predicted curves 
are consistent with the H.E.S.S. data despite the fact that we did not 
use it in our fit procedure.  Note that the background fluxes are very 
uncertain above $\approx 1~{\rm TeV}$, so that the precise comparison 
with the H.E.S.S. data may be misleading; for example, the background 
spectrum adopted here seems a bit too hard at the highest energies 
to be consistent with the H.E.S.S. data.

To summarize, we find that the electron/positron fluxes observed by 
PAMELA and FERMI are very well explained by dark matter decay in our 
scenario.  The mass of the axion-like state can take almost any value 
in the wide range considered in Section~\ref{subsec:axion}, except 
for a small window at $\simeq 1.9~\mbox{--}~3.6~{\rm GeV}$ where $a$ 
decays hadronically.  More specifically, we find that the regions
\begin{equation}
\begin{array}{rlll}
  \mbox{(i)} \,\,\,   & a \rightarrow e^+e^- 
    & (2 m_e < m_a < 2 m_\mu) \quad 
    & f_a \approx O(100~{\rm TeV}), \\
  \mbox{(ii)} \,\,\,  & a \rightarrow \mu^+\mu^- 
    & (2 m_\mu < m_a \simlt m_K - m_\pi) \quad 
    & f_a \approx O(10~\mbox{--}~100~{\rm TeV}), \\
  \mbox{(ii)} \,\,\,  & a \rightarrow \mu^+\mu^- 
    & (m_K - m_\pi < m_a \simlt 800~{\rm MeV}) \quad 
    & f_a \approx O(1~\mbox{--}~100~{\rm TeV}), \\
  \mbox{(iii)} \,\,\, & a \rightarrow \pi^+\pi^-\pi^0 
    & (800~{\rm MeV} \simlt m_a < 2 m_p) \quad 
    & f_a \approx O(1~\mbox{--}~100~{\rm TeV}), \\
  \mbox{(iv)} \,\,\,  & a \rightarrow \tau^+\tau^- 
    & (2 m_\tau < m_a < 2 m_b) \quad 
    & f_a \approx O(1~\mbox{--}~100~{\rm TeV}),
\end{array}
\label{eq:axion-final}
\end{equation}
can all explain the cosmic ray $e^\pm$ data, without conflicting the 
bounds discussed in Section~\ref{subsec:axi-const}.

\subsection{Diffuse gamma ray signals at FERMI}
\label{subsec:gamma}

An immediate consequence of the framework described here is that the 
decay of dark matter will provide a source of $\gamma$ rays throughout 
the dark matter halo, and extending up to energies around a few TeV. 
If the axion-like state decays into $\tau^+\tau^-$ or $\pi^+\pi^-\pi^0$, 
photons are produced directly by the decay of $\pi^0$s, while for 
$e^+e^-$ and $\mu^+\mu^-$ modes there is a much smaller but significant 
source of $\gamma$ rays from final state radiation (FSR) and inverse 
Compton scattering (ICS).  Although the greatest fluxes would originate 
from the galactic center, where number densities are highest, the best 
direction to look for them is away from the galactic plane, where 
the background is much smaller and the signal still large.  The 
FERMI experiment will measure $\gamma$-ray fluxes over the entire 
sky at energies up to several hundred GeV, which has the potential 
to resolve spectral features caused by dark matter 
decays~\cite{Baltz:2008wd,Regis:2009md}.

\begin{figure}
  \center{\includegraphics[scale=1.0]{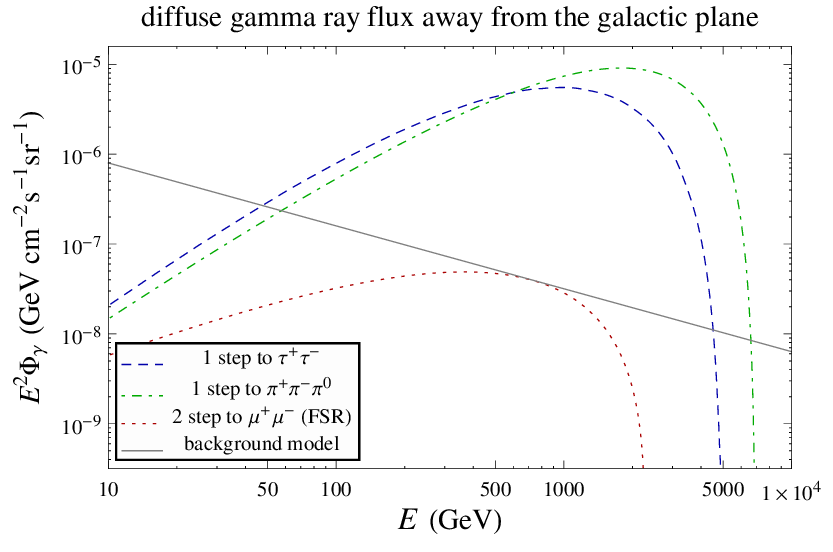}}
\caption{The high energy diffuse $\gamma$-ray spectrum away from the 
 galactic plane, averaged over galactic latitudes above $10^\circ$ 
 assuming an NFW profile.  Shown are the best fit parameters for 
 $1$-step to $\tau^+\tau^-$ (dashed), $1$-step to $\pi^+\pi^-\pi^0$ 
 (dot-dashed), and $2$-step to $\mu^+\mu^-$ (dotted) dark matter 
 decay modes.  The $\gamma$-rays are due to $\pi^0$ decay in the 
 case of $1$-step to $\tau^+\tau^-$ and $\pi^+\pi^-\pi^0$, and to 
 FSR in the case of $2$-step to $\mu^+\mu^-$.  The $\tau^+\tau^-$ 
 and $\pi^+\pi^-\pi^0$ modes produce a bump in the flux clearly 
 distinguishable from the background.}
\label{fig:gamma}
\end{figure}
In Figure~\ref{fig:gamma} we plot the contributions to the diffuse 
$\gamma$-ray fluxes for three illustrative decay modes: $1$-step 
to $\tau^+\tau^-$, $1$-step to $\pi^+\pi^-\pi^0$, and $2$-step to 
$\mu^+\mu^-$.  The first two cases are representative of any cascade 
to taus or pions, while the $\mu^+\mu^-$ curve illustrates the 
much lower flux for a decay mode with only FSR photons.  Assuming 
an NFW profile, we average over all galactic latitudes greater 
than $10^\circ$ from the galactic plane, representative of 
a diffuse $\gamma$-ray measurement by FERMI.  The astrophysical 
background is modeled by a power law flux, $\propto E^{-2.7}$, 
taken from~\cite{Bergstrom:1997fj}.  For the $\mu^+\mu^-$ mode 
the flux shown is solely due to FSR, and we assume an $a$ mass 
of $600~{\rm MeV}$.  We have not shown the contribution from ICS, 
which, like FSR, is subdominant to directly produce photons for 
$\tau^+\tau^-$ and $\pi^+\pi^-\pi^0$ modes.  At its peak, the ICS 
flux-component is expected to be comparable to the background, 
with a spectral shape somewhat different to that from 
FSR~\cite{Meade:2009iu,Regis:2009md}, and could be significant 
for $e^+e^-$ and $\mu^+\mu^-$ modes.

We find that signals from dark matter in our scenario can be seen 
in the diffuse $\gamma$-ray data.  While the background is very 
uncertain, it is expected to be smooth compared to the strong, peaked 
fluxes seen for the $\tau^+\tau^-$ and $\pi^+\pi^-\pi^0$ decay modes, 
which should result in a clearly visible bump in the spectrum.  The 
weaker signals from $e^+e^-$ and $\mu^+\mu^-$ modes may also be seen. 
Since astrophysical sources are not expected to produce $\gamma$-ray 
fluxes with such prominent spectral features, measurements of diffuse 
$\gamma$-rays may serve to distinguish dark matter as the source of 
the PAMELA and FERMI excesses.  Additionally, the shape and the strength 
of the $\gamma$-ray spectrum would convey information about the mass 
and decay channels of the dark matter.  While the absence of an excess 
in upcoming FERMI data would not exclude our scenario, especially 
for the $e^+e^-$ and $\mu^+\mu^-$ cases, a positive result would be 
a striking signature.

\section{Collider Signatures}
\label{sec:collider}

If the PAMELA and FERMI data is indeed indicative of dark matter with 
an $O(10~{\rm TeV})$ mass, then direct production of dark matter is not 
possible at current collider experiments.  However, the light axion-like 
states, which play a crucial role in determining the cosmic ray spectra, 
are kinematically accessible.  We briefly describe some of the collider 
signatures for the axion-like state $a$.

The discovery potential for the axion-like state depends strongly on 
its mass and decay constant.  Since we are considering decay constants 
in the range $f_a \approx O(1~\mbox{--}~100~{\rm TeV})$, the coupling 
of $a$ to the standard model is small, so discovery relies on the 
axion-like state having a clean decay mode.  For $2 m_e < m_a < 2 m_\mu$, 
the decay length is generically greater than a kilometer, so any axion-like 
state produced in a collision will decay outside of the detector.  In mass 
range $2 m_\mu < m_a \simlt 800~{\rm MeV}$, $a$ decays into $\mu^+\mu^-$ 
with a possible displaced vertex, offering a promising discovery 
channel.  For $2 m_\tau < m_a < 2 m_b$, $a$ decays promptly into 
taus, but neither leptonic taus nor hadronic taus are particularly 
clean channels. Similarly, there are large hadronic backgrounds 
to $a \rightarrow \pi^+\pi^-\pi^0$.  Interestingly, while the 
$\tau^+\tau^-$ and $\pi^+\pi^-\pi^0$ channels are challenging in 
the collider context, they are precisely the ones that give the most 
striking diffuse $\gamma$-ray signal as seen in Section~\ref{subsec:gamma}. 
In contrast, the $\mu^+\mu^-$ final state yields less dramatic diffuse 
$\gamma$-ray flux, but relatively clean collider signatures.

Direct production of $a$'s at the LHC was considered in 
Ref.~\cite{Goh:2008xz}, where the $a$ is produced in association 
with a hard ($p_T > 100~{\rm GeV}$) jet via the gluon-gluon-$a$ coupling. 
Since the $a$ is boosted from the production, the resulting muon 
tracks from $a$ decay have a mrad opening angle.  If the decay constant 
$f_a \simgt (3~\mbox{--}~10)~{\rm TeV}$, then the axion-like state lives 
sufficiently long that the decay $a \rightarrow \mu^+\mu^-$ happens 
with an $O({\rm cm})$ displaced vertex.  As long as the decay constant 
$f_a \simlt (15~\mbox{--}~30)~{\rm TeV}$, then the production rate is 
sufficiently large to compete with the $B^0 \rightarrow D^\pm\mu^\mp\nu 
\rightarrow \mu^+\mu^-\nu\nu$ background.  Therefore, direct production 
of $a$ is a promising possibility for $f_a \sim 10~{\rm TeV}$.

An alternative production mechanism for $a$'s is via the Higgs boson. 
For the $R$ and PQ symmetries considered in Section~\ref{subsec:axion}, 
the axion-like state has couplings to the Higgs fields, allowing the Higgs 
boson to decay into two axion-like states.  For simplicity, we will focus 
on the decoupling limit as in Eq.~(\ref{eq:a-coupling-h_dec}).  The Higgs 
decay width into two $a$'s is then given by
\begin{equation}
  \Gamma(h \rightarrow aa) = \frac{c_1^2}{64\pi} \frac{v^2 m_h^2}{f_a^4},
\label{eq:h-aa}
\end{equation}
where $c_1 = (1/4)\sin^2\!2\beta$, and $m_h$ is the Higgs boson mass. 
For a light Higgs boson, the dominant decay mode is into $b\bar{b}$ 
with (taking $m_b/m_h \rightarrow 0$)
\begin{equation}
  \Gamma(h \rightarrow b\bar{b}) = \frac{3}{16\pi} m_h 
    \left(\frac{m_b}{v}\right)^2.
\label{eq:h-bb}
\end{equation}
Therefore, the branching ratio into two axion-like states is
\begin{equation}
  {\rm Br}(h \rightarrow aa) \simeq 
    \frac{c_1^2}{12} \frac{m_h^2 v^4}{m_b^2 f_a^4} 
  \simeq 1.3 \times 10^{-5} 
    \left(\frac{m_h}{120~{\rm GeV}} \right)^2 
    \left(\frac{10~{\rm TeV}}{f_a \tan\beta} \right)^4,
\label{eq:br-haa}
\end{equation}
where we have used the large $\tan\beta$ approximation, $\sin\!2\beta 
\approx 2/\tan\beta$, in the last equation.  We see that the branching 
fraction for $h \rightarrow aa$ depends strongly on the decay constant 
as $f_a^{-4}$, and a sizable branching fraction can be obtained as 
$f_a$ approaches $1~{\rm TeV}$.  Since the decay constant is naturally 
expected to be somewhat ($\approx 4\pi$) smaller than the dynamical 
scale $\Lambda$ as we saw in Eq.~(\ref{eq:f_a}), this makes observation 
of the $h \rightarrow aa \rightarrow 4\mu$ signal at the LHC an 
interesting possibility~\cite{Nomura:2008ru}.  The recent analysis 
of Ref.~\cite{Lisanti:2009uy} focused on the $h \rightarrow aa 
\rightarrow 2\mu 2\tau$ channel in the case of $m_a \sim 5~{\rm GeV}$. 
They find that generic event selection has an efficiency around 
$10\%$.  Assuming a similar efficiency for $h \rightarrow aa \rightarrow 
4\mu$ events, and taking the Higgs production at the LHC of $\sim 
50~{\rm pb}$, at least 10 events could be seen with $300~{\rm fb}^{-1}$ 
of data for $f_a \tan\beta \simlt 10~{\rm TeV}$.  In this way, direct 
$a$ production (for larger values of $f_a$) and $a$ production through 
the Higgs boson (for smaller values of $f_a$) are complementary. 
For larger $m_a$, the $h \rightarrow aa \rightarrow 6\pi$ and 
$h \rightarrow aa \rightarrow 4\tau$ channels may be visible if 
the background can be controlled.  Note that the recent D\O\ analysis 
in Ref.~\cite{Abazov:2009yi} already gives the constraint $f_a \tan\beta 
\simgt 2~{\rm TeV}$ for $a \rightarrow \mu^+\mu^-$.

There has also been recent interest in looking for light bosons 
in low-energy high-luminosity lepton colliders~\cite{Batell:2009yf,%
Reece:2009un} as well as in fixed-target experiments~\cite{Reece:2009un}. 
From Eq.~(\ref{eq:a-coupling-f}), the coupling of the axion-like state 
to electrons is proportional to $m_e/f_a$, which is smaller than 
$10^{-6}$ for $f_a > 1~{\rm TeV}$, making the process $e^+ e^- 
\rightarrow \gamma a$ beyond the reach of current lepton colliders. 
The feasibility of a fixed-target experiment to discover $a$ depends 
on the $a$ lifetime.  If $a$ decays promptly, then one must contend 
with a huge standard model background of prompt charged particle 
production, though Ref.~\cite{Reece:2009un} suggests that a coupling 
as small as $10^{-6}$ might be testable if $a$ decays into $\mu^+\mu^-$. 
If $a$ decays with a displaced vertex, then it could be discovered 
in traditional electron or proton beam dump experiments.  Given 
the bounds from Figure~\ref{fig:ma-fa} and the lifetimes in 
Eqs.~(\ref{eq:ct-ee},~\ref{eq:ct-mumu},~\ref{eq:ct-tautau}), 
the possible values of $c\tau$ spans a huge range from tens of 
kilometers to less than a nanometer.  If $a \rightarrow \mu^+\mu^-$, 
then $c\tau$ is plausibly in the millimeter to centimeter range, 
and could likely be discovered in an upgraded version of the 
experiment from Ref.~\cite{Bergsma:1985qz}.

\section{Discussion and Conclusions}
\label{sec:discuss}

The origin of dark matter is one of the greatest mysteries in particle 
physics and cosmology.  From the theoretical point of view, attentions 
have been focused on the WIMP paradigm: dark matter has mass $m_{\rm DM} 
\approx O(100~{\rm GeV}~\mbox{--}~1~{\rm TeV})$ and couplings of weak 
interaction strength $g \approx O(1)$, leading to an annihilation cross 
section that gives the correct thermal relic abundance, $\Omega_{\rm DM} 
\simeq 0.2$.  However, since the annihilation cross section depends on 
the combination $g^2/m_{\rm DM}$, it should be equally convincing to 
consider the case where $m_{\rm DM}$ is heavier than the weak scale as 
long as the coupling $g$ is larger.  Such a situation arises naturally 
if dark matter is a composite state of some strong interaction, since 
the typical coupling $g$ is expected to be larger than order unity and 
$m_{\rm DM}$ can be $O(10~{\rm TeV})$.

Suppose that dark matter indeed arises from some strongly interacting 
sector.  Then its stability may be the result of compositeness, and 
not of some exact symmetry imposed on the theory.  This is precisely 
analogous to the case of the proton in the standard model embedded in 
some unified theory.  If the proton (and pions) were elementary, it 
would immediately decay into $e^+$ and $\pi^0$ through a Lagrangian 
term ${\cal L} \sim p e \pi^0$.  Since it is composite, however, 
the leading operator causing proton decay is already dimension six, 
${\cal L} \sim q q q l/M_*^2$, and the resulting lifetime is of 
order $10^{36}~{\rm years}$ for $M_*$ of order the unification scale. 
In the case of composite dark matter, dimension six operators yield 
a lifetime of order $10^{25}~{\rm sec}$ for $m_{\rm DM} \approx 
10~{\rm TeV}$ and $M_* \approx 10^{17}~{\rm GeV}$.  Decay of galactic 
dark matter could then have currently observable consequences.

The story just described implies that there is new strongly coupled 
physics beyond the weak scale, at $\approx O(10~\mbox{--}~100~{\rm TeV})$. 
Interestingly, we already know an attractive framework where such a 
situation occurs---weak scale supersymmetry with low energy dynamical 
supersymmetry breaking.  Since the superparticle masses in this framework 
arise at loop level, the scale of the strong sector is naturally larger 
than the weak scale by a one-loop factor $\approx 16\pi^2$.   This 
picture is very much consistent with what is implied by the LEP precision 
electroweak data, namely that physics at the weak scale itself is 
weakly coupled.  Yet such a setup can still explain the large hierarchy 
between the weak and Planck scales in, arguably, the simplest manner 
via dimensional transmutation.

It is interesting that this picture of dark matter arises precisely 
in the scenario where conventional LSP dark matter is lost---the 
LSP is now the very light gravitino.  In fact, a gravitino with 
mass $m_{3/2} \simlt O(10~{\rm eV})$, implied by $\Lambda \approx 
O(10~\mbox{--}~100~{\rm TeV})$, avoids various cosmological difficulties 
faced by other supersymmetry breaking scenarios.  This allows us to 
consider the ``standard'' cosmological paradigm, based on inflation 
at a very early epoch with subsequent baryogenesis at high energies, 
consistently with supersymmetry.  The standard virtues of weak scale 
supersymmetry, such as the stability of the weak scale and gauge 
coupling unification, are all preserved.

While composite dark matter in low-scale supersymmetry breaking already 
offers a consistent picture, this may not be the end of the story. 
Since dark matter is a part of a strongly interacting sector, it is 
possible that it feels additional dynamical effects.  In particular, 
it is quite plausible that the sector spontaneously breaks an 
accidental global symmetry, leading to a light pseudo Nambu-Goldstone 
boson---again as in QCD.  This raises the possibility that dark matter 
decays mainly into these light states (possibly through some other 
state), which then decay into standard model particles.  In supersymmetric 
theories there is a natural candidate for such symmetry, an $R$ symmetry, 
whose existence is suggested by a certain genericity argument associated 
with supersymmetry breaking.  The mass of these light states can easily 
be in the ${\rm MeV}$ to $10~{\rm GeV}$ range if explicit breaking 
arises from dimension five operators.   Except in the mass range 
$m_a \simeq 1.9~\mbox{--}~3.6~{\rm GeV}$, the decay products of 
the light states are naturally ``leptonic'' (specifically, $e^+e^-$, 
$\mu^+\mu^-$, $\pi^+\pi^-\pi^0$, or $\tau^+\tau^-$), with little 
nucleonic activity.

An illustrative model of this class was given in Section~\ref{sec:illust}, 
and the required symmetry structure for the supersymmetry breaking 
sector was summarized in Table~\ref{tab:symmetry}.  Remarkably, the 
properties we have just obtained (with a little adjustment of parameters), 
are precisely what is needed to explain the recent cosmic ray data 
through dark matter physics:
\begin{itemize}
\item
Dark matter mass is of $O(10~{\rm TeV})$,
\item
Dark matter lifetime is of $O(10^{26}~{\rm sec})$,
\item
Dark matter decays through (long) cascades,
\item
Dark matter decay final states are leptonic.
\end{itemize}
As we saw in Section~\ref{sec:astro}, various mysterious features in the 
data---an unexpected rise of the positron fraction between $\approx 10$ 
and $100~{\rm GeV}$ in the PAMELA data, nonobservation of any anomaly 
in the PAMELA antiproton data, and a broad excess of the combined 
$e^\pm$ flux in the FERMI data---are all beautifully explained by 
the properties of dark matter discussed in this paper.  The resulting 
$e^\pm$ spectra are also consistent with the recent H.E.S.S. result. 
The success is quite remarkable, especially in view of the fact 
that the data is difficult to explain in terms of conventional WIMP 
annihilation---the mass scale suggested does not seem natural, the 
observed rate requires a large boost factor, and typical WIMP annihilation 
produces more antiprotons than indicated by PAMELA.

While dark matter with mass of order $10~{\rm TeV}$ will not be produced 
at the LHC, the present scenario still has potential collider signatures. 
Since the light states generically have interactions with standard model 
gauge bosons, they may be produced at the LHC, and the leptonic decay 
of the light state could lead to visible signatures, especially if 
the dominant decay is into muons.  The Higgs boson may also decay into 
the light states, producing a clean four lepton final state.  Future 
astrophysical observations could also probe this scenario.  For example, 
measurements of diffuse $\gamma$-ray could discriminate $e$/$\mu$ final 
states from $\tau$/$\pi$ final states.  Moreover, the present framework 
may explain the discrepancy between the measured ${}^7{\rm Li}$ abundance 
and the standard big-bang nucleosynthesis prediction by dimension 
five decay of some of the states in the supersymmetry breaking sector, 
e.g.\ (composite) messenger fields.  Analysis of all these experimental 
data could provide important information about the structure of the 
supersymmetry breaking sector.

We finally mention that while in this paper we focused on the case where 
dark matter arises from a strongly interacting supersymmetry breaking 
sector and decays through light states, some of our results apply in 
wider contexts.  For example, quasi-stable dark matter in the supersymmetry 
breaking sector may decay directly into the SSM sector particles through 
dimension six operators.  For example, quasi-stable mesons can decay 
into SSM (s)leptons $L_{\rm SSM}$ through the K\"{a}hler potential 
operators $K \sim Q^\dagger_i Q^j L_{\rm SSM}^\dagger L_{\rm SSM}/M_*^2 
+ \bar{Q}^{\dagger\bar{\imath}} \bar{Q}_{\bar{\jmath}} L_{\rm SSM}^\dagger 
L_{\rm SSM}/M_*^2$.  In these cases, the final states of the decay are 
determined by a combination of gravitational scale physics and TeV-scale 
superparticle spectra, not through cascades associated with light states. 
However, all other points regarding compositeness, long lifetime, 
relatively large mass, and thermal abundance still persist.  The dynamics 
of dark matter discussed here can also be applied in non-supersymmetric 
theories---all we need is some strong dynamics at $\approx 
O(10~\mbox{--}~100~{\rm TeV})$ satisfying the properties of 
Table~\ref{tab:symmetry}.  Examples of such theories may include 
ones in which the Higgs fields arise as pseudo Nambu-Goldstone 
bosons of some strong dynamics~\cite{Kaplan:1983fs}.

As the LHC starts running this year, we anticipate great discoveries. 
A possible connection between weak scale physics and dark matter is 
one of the major themes to be explored at the high energy frontier. 
The standard expectation is to study the properties of a dark matter 
particle by producing it at the LHC.  Although the scenario presented in 
this paper does not allow this, important physics associated with the 
dark sector may still be probed.  With many new particle physics and 
astrophysics observations on the horizon, the next decade will certainly 
be exciting for potentially understanding the origin and properties 
of dark matter.  While nature may not show us the ``standard'' WIMP 
dark matter story, it may still reveal a beautiful connection between 
dark matter and the weak scale through hidden sector dynamics.

\section*{Acknowledgments}

This work was supported in part by the Director, Office of Science, 
Office of High Energy and Nuclear Physics, of the US Department of 
Energy under Contract DE-AC02-05CH11231, and in part by the National 
Science Foundation under grant PHY-0457315.  The work of Y.N. was 
supported in part by the National Science Foundation under grant 
PHY-0555661 and the Alfred P. Sloan Foundation.  J.T. is supported 
by the Miller Institute for Basic Research in Science.

\end{document}